\documentclass[aps,prb,preprint,superscriptaddress,superscriptreference]{revtex4-1}
\usepackage{amsmath,bbold,bm,amssymb,scalerel,mathtools}
\usepackage{graphicx}
\usepackage{color}
\usepackage{hyperref}
\usepackage{enumitem}
\usepackage{algorithm,algpseudocode}
\usepackage{multirow}
\usepackage{colortbl,booktabs}
\usepackage{placeins}
    \usepackage{url}

\AtBeginDocument{\nocite{achemso-control}}
\newcommand*{\citen}{}% generate error, if `\citen` is already in use
\DeclareRobustCommand*{\citen}[1]{%
	\begingroup
	\romannumeral-`\x % remove space at the beginning of \setcitestyle
	\setcitestyle{numbers}%
	\cite{#1}%
	\endgroup
}

\newcommand\equalhat{%
	\let\savearraystretch\arraystretch
	\renewcommand\arraystretch{0.3}
	\begin{array}{c}
		\stretchto{
			\scalerel*[\widthof{=}]{\wedge}
			{\rule{1ex}{3ex}}%
		}{0.5ex}\\ 
		=%
	\end{array}
	\let\arraystretch\savearraystretch
}
\newcommand{\myrowcolour}{\rowcolor[gray]{0.925}}
\newcommand{\gj}[1]{\textcolor{black}{#1}}

\usepackage[usenames,dvipsnames]{xcolor}
\usepackage[normalem]{ulem}

\begin{document}

\title{Scaling equations for mode-coupling theories with multiple decay channels}

\author{Gerhard Jung}
\email{gerhard.jung@uibk.ac.at}
\affiliation{Institut f\"ur Theoretische Physik,  Universit\"at Innsbruck, Technikerstra{\ss}e 21A, 6020 Innsbruck, Austria}

\author{Thomas Voigtmann}
\email{thomas.voigtmann@hhu.de}
\affiliation{Institut f{\"u}r Materialphysik im Weltraum, Deutsches Zentrum f{\"u}r Luft- und Raumfahrt (DLR), 51170 K{\"o}ln, Germany}
\affiliation{Institut f{\"u}r Physik, Heinrich-Heine-Universit\"at, Universit{\"a}tsstra{\ss}e 1, 40225 D{\"u}sseldorf, Germany}

\author{Thomas Franosch}
\email{thomas.franosch@uibk.ac.at}
\affiliation{Institut f\"ur Theoretische Physik, Universit\"at Innsbruck, Technikerstra{\ss}e 21A, 6020 Innsbruck, Austria}

\begin{abstract}
Multiple relaxation channels often arise in the dynamics of liquids where the momentum current associated to the particle-conservation law splits into distinct contributions. Examples are strongly confined liquids for which the currents in lateral and longitudinal direction to the walls are very different, or fluids of nonspherical particles with distinct relaxation patterns for translational and rotational degrees of freedom. Here, we perform an asymptotic analysis of the slow structural relaxation close to kinetic arrest 
as described by  mode-coupling theory (MCT) with several relaxation channels. Compared to standard MCT, the presence of multiple relaxation channels significantly changes the structure of the underlying equations of motion and leads to additional, non-trivial terms in the asymptotic solution. We show that the solution can be rescaled, and thus prove that the well-known $ \beta $-scaling equation of MCT remains valid  even in the presence of multiple relaxation channels. The asymptotic treatment is validated using a novel schematic model. We demonstrate that the numerical solution of this schematic model can indeed be described by the derived asymptotic scaling laws close to kinetic arrest. Additionally, clear traces of the existence of two distinct decay channels are found in the low-frequency susceptibility spectrum, suggesting that clear footprints of the additional relaxation channels can in principle be detected in simulations or experiments of confined or molecular liquids.

\end{abstract}

\maketitle

\section{Introduction}

The mode-coupling theory of the glass transition (MCT) has been conceived and elaborated to  describe the rapid slowing-down of the structural relaxation in (supercooled) liquids and has become a well-established and very 
successful approach for the theoretical analysis of the slow dynamics in a variety of systems. In particular, mode-coupling theory and its extensions  include now simple liquids~\cite{Bengtzelius_1984,Gotze_1992,Gotze2009}, colloidal mixtures \cite{Gotze1987,Bosse1987,Barrat_1990,Szamel:PRA_44:1991,Kob1995,Voigtmann2003}, non-spherical particles (rigid molecules) \cite{Franosch1997a,Goetze2000c,Schilling:PRE_56_1997,Kaemmerer:PRE_56_1997,Kaemmerer:PRE_58_1998,Kaemmerer:PRE_58_1998b,Fabbian1999,Winkler2000,Chong1998,Chong2000}, 
random host structures \cite{Krakoviack:PRL_94_2005,Krakoviack:PRE_84_2011} and confined systems \cite{Lang2010, Lang2012,Mandal2014, 
Mandal2017a}, 
sheared colloidal liquids~\cite{Fuchs2002,Brader2007,Brader2009,Fuchs2009},  driven granular fluids \cite{Sperl:PhysRevLett.104.225701,Sperl:PhysRevE.87.022207}
active microrheology \cite{Gazuz2009,Gruber2016,Gruber2019_2,Senbil2019} and active particles \cite{Nandi2017,Liluashvili2017,Janssen_2019}.

A central prediction of MCT for the dynamics of simple liquids is the universal relaxation behavior close to the ideal glass transition. 
In precisely defined asymptotic limits, 
MCT predicts a two-time fractal for any time-correlation function of observables in the vicinity of a critical plateau value. The \emph{critical law}  emerges directly as long-time relaxation towards this plateau as the glass-transition singularity is reached in the non-equilibrium state diagram. Already at close distances below the singularity this critical law becomes apparent, yet is superseded by another power law referred to as von Schweidler law characterizing the initial decay from the critical plateau~\cite{Gotze_1990,Gotze_1992,Gotze2009, Franosch1997}. The two time fractals are intimately related and give rise to the first scaling law ($\beta$-scaling) of the theory in terms of a  universal scaling function. In particular, the exponents are related by a non-algebraic relation and the factorization theorem holds, i.e., the dynamical correlation functions asymptotically close to the transition are all governed by the same scaling function, where observable-dependent prefactors only enter in a critical amplitude. The governing equations for the scaling function  have been derived first  by G{\"o}tze \cite{Gotze_1990}
and later extended also for the leading corrections \cite{Franosch1997} for scalar quantities. The case of matrix-valued time correlation functions \cite{Voigtmann2019} and the scaling function for generalized mode-coupling theory \cite{Szamel:PhysRevLett.90.228301,JansenReview2018,Janssen_2019_GMCT} has been elaborated only recently. 

A second scaling law ($\alpha$-scaling) is predicted for the ultimate long-time relaxation characterizing the decay of time-correlation functions below the glass-transition singularity~\cite{Gotze2009}. This long-time dynamics
 is stretched, it extends over many decades in time and is the theoretical explanation for the phenomenological time-temperature superposition principle and the empirical  Kohlrausch function.    Both scaling laws have been observed in a series of  photon-correlation experiments on colloids \cite{VanMegen1993,VanMegen1994}, depolarized light-scattering \cite{Li:PRA_45_1992,Li:PRA_46_1992,Franosch1997,Singh1998} or dielectric spectroscopy \cite{Lunkenheimer:PRL_77_1996,Schneider1999,Goetze2000}  on supercooled liquids  
as well as in  computer simulations of simple mixtures \cite{Kob1994,Horbach2008}, water \cite{Gallo:PRL_76_1996,Sciortino:PRE_54_1996,Sciortino:PRE_56_1997} or silica glasses \cite{Sciortino:PRL_86_2001,Horbach1998,Voigtmann2006}.   
 
 Many of the above mentioned extensions to MCT, viz. rigid molecules, confined systems, microrheology and active particles, have in common that the relaxation of \gj{identical} particles in a system is governed by distinct decay channels for different degrees of freedom. For molecules and active particles these are the rotational and translational degrees of freedom\cite{Franosch1997a,Chong1998,Fabbian1999,Chong2000,Winkler2000,Nandi2017,Liluashvili2017}, in confined systems the directions parallel and perpendicular to the wall\cite{Lang2010, Lang2012,Mandal2014, Mandal2017a} and in active microrheology the directions along and perpendicular to the active force~\cite{Gruber2016,Gruber2019_2}. The emergence of multiple decay channels changes the overall mathematical structure of the underlying equations of motion and it is not obvious whether the $\beta $-scaling equation also translates to those situations.

For the case of a single molecule dissolved in a simple liquid, the validity of the factorization theorem has been proved \footnote{See Ref.~[\citen{Franosch1997a}] and A.~P.~Singh, \emph{PhD thesis}, Technische Universit{\"a}t M{\"u}nchen; 1998}, stating that the critical dynamics can be factorized into a (matrix-valued) critical amplitude and a scalar, time-dependent function, called the $ \beta$-correlator. We will encounter this theorem later as a result of the first-order expansion. 
	For the collective dynamics in situations where parallel relaxation 
	plays a role, asymptotic expansions have been used to determine critical amplitudes \cite{Winkler2000,Rinaldi2001} (based on the factorization theorem), yet, no 
	theoretical justification of the $ \beta $-scaling equation was presented.
	In fact, a recent asymptotic analysis for a particle with constant pulling force in a simple liquid (active microrheology) \cite{Gruber2019_2} showed
	that in this special case {a mixed continuous/discontinuous transition} could be observed but no two-step-relaxation scenario.
	A rigorous calculations is thus required to establish the $\beta$-scaling regime for the recent extensions of MCT mentioned above.

The purpose of this work is to derive the $ \beta $-scaling equation for mode-coupling theories with multiple decay channels from asymptotic analysis. First, we derive in Sec.~\ref{sec:analysis} 
the evolution equations for the structural relaxation and show that new terms arise in the asymptotic expansion due to the existence of distinct decay channels. Using the scale-invariance of the scaling equation 
we suggest a straightforward rescaling of the governing dynamical quantities to recover the original form of the $\beta$-scaling equation. To validate the calculations we introduce in Sec.~\ref{sec:schematic} 
a schematic model with two decay channels, based on the Bosse-Krieger model \cite{Krieger1987}. We perform a detailed analysis of the critical dynamics of this schematic model and evaluate the results using the previously derived asymptotic scaling laws. We summarize and conclude in Sec.~\ref{sec:conclusion}. 

\section{Asymptotic analysis and scaling equations}
\label{sec:analysis}

In the first part of this chapter, we will recapitulate the mode-coupling equations for systems with multiple decay channels as they were presented in Ref.~[\citen{Lang_2013}]. We will focus on identifying the introduced quantities on the example of a simple liquid in confined geometry \gj{and subsequently discuss how the formalism can be generalized, for example to describe molecular liquids\cite{Schilling:PRE_56_1997,Kaemmerer:PRE_58_1998,Kaemmerer:PRE_58_1998b,Fabbian1999}}. We will then write down the equations for structural relaxation close to the glass transition and perform an asymptotic expansion to derive the factorization theorem and the $ \beta $-scaling equation. For readers who are not interested in the technical details, we discuss the most important results of the asymptotic analysis in Sec.~\ref{sec:discussion}.

\subsection{\gj{Evolution equations of a simple liquid in confinement}}

\gj{Let us consider a system of $ N $ identical particles, confined in z-direction between two parallel, hard walls \cite{Lang2012}. The system can be described by the microscopic particle density,
\begin{equation}
\rho(\bm{r},z,t) = \sum_{i=1}^{N} \delta \left[\bm{r} - \bm{r}_i(t)\right] \delta\left[ z-z_i(t) \right],
\end{equation}
with $ \bm{r} = (x,y) $. Due to translational symmetry in the directions parallel to the walls, we find that the averaged density $ \left\langle  \rho (\bm{r},z,t)\right\rangle = n(z) $ only depends on the z-coordinate. To characterize the dynamics of the microscopic system we introduce the Van Hove correlation function,
\begin{equation}
G(|\bm{r}-\bm{r}'|,z,z',t) = n_0^{-1} \left\langle  \delta \rho (\bm{r},z,t)\delta \rho (\bm{r}',z',t)\right\rangle,
\end{equation}
as the time-correlation function of the microscopic density fluctuations, $ \delta \rho (\bm{r},z,t) = \rho (\bm{r},z,t) - n(z) $, normalized by the area density $ n_0 $. It is now natural and numerically convenient to introduce density modes as the Fourier transform of the particle density, $ \rho_\mu(\bm{q},t)=\sum_{i=1}^N \exp[{\rm i}Q_\mu z_i(t)]\exp[{\rm i}\bm{q}\cdot\bm {r}_i(t)]  $. The modes perpendicular to the confining direction are discrete,  
$  Q_\mu = 2 \pi \mu / L $, in contrast to the continuous wave vectors $ \bm{q} $ in the lateral direction. Here, $ L $ is the accessible slit width. From this we can define the generalized intermediate scattering function (ISF),
\begin{equation}\label{key}
S_{\mu\nu}(q,t)= \frac{1}{N} \left \langle  \rho_\mu(\bm{q},t)^* \rho_\nu(\bm{q},0) \right \rangle.
\end{equation}
By Fourier transformation one immediately finds that the ISF is nothing but the Fourier transform of the above introduced Van Hove correlation function \cite{ Lang2012},
\begin{align}
S_{\mu \nu}(q,t) = \int_{-L/2}^{L/2} \text{d}z\int_{-L/2}^{L/2} \text{d}z' \int_{A}^{} \text{d}(\bm{r}-\bm{r}')G(|\bm{r}-\bm{r}'|,z,z',t)
 \exp \left[-\textrm{i}(Q_\mu z - Q_\nu z')  \right] e^{-\textrm{i}\bm{q}\cdot(\bm{r}-\bm{r}')}.
\end{align} }

Choosing the density modes $ \left\{ \rho_\mu(\bm{q},t) \right\} $ as set of distinguished variables,  the Zwanzig-Mori projection operator formalism\cite{Zwanzig2001,Gotze2009,Hansen:Theory_of_Simple_Liquids} yields the equations of motion for the intermediate scattering function,
\begin{equation}\label{eq:eom1}
\dot{\mathbf{S}}(t)+\int_0^t \mathbf{K}(t-t')\mathbf{S}^{-1}\mathbf{S}(t') \text{d}t' =0.
\end{equation}
Here, we have introduced the matrix notation $ \left[ \mathbf{S}(t) \right]_{\mu \nu} = S_{\mu\nu}(q,t) $ and the generalized structure factor $ \mathbf{S} = \mathbf{S}(t=0) $. Occasionally the dependence on the wavevector $\bm{q}$ will be suprressed in cases where it serves merely as a parameter. 
The \emph{a priori} unknown memory kernel $ \left[\mathbf{K}(t)\right]_{\mu\nu} =  K_{\mu\nu}(q,t) $ defines the non-Markovian dynamics of the ISF and 
corresponds formally to the time correlation function of the time derivatives of the  density fluctuations, $ \dot{\rho}_\mu(\bm{q}) $ albeit with projected dynamics.  
More precisely, it is the correlation function of 
$ \dot{\rho}_\mu(\bm{q},0) $ 
with $ e^{(1-\mathcal{P})\mathcal{L}t  }  \dot{\rho}_\mu(\bm{q},0) $, 
where $ \mathcal{L} $ is the standard Liouville operator and 
$ \mathcal{P} $ the Zwanzig-Mori projection operator \cite{Zwanzig2001,Gotze2009}.

To derive an expression for the memory kernel, we use the fact that the density modes fulfill the continuity equation,
\begin{equation}\label{eq:continuity}
\dot{\rho}_\mu(\bm{q},t) = { \rm i} \sum_{\alpha=1}^m q_\mu^\alpha j^\alpha_\mu(\bm{q},t),
\end{equation}
where the superscript $\alpha$ is referred to as channel index and  the $ q_\mu^\alpha $ define the fundamental couplings of the current channel $ j^\alpha_\mu(\bm{q},t) $ to the density modes.  
Since we are interested in systems with multiple relaxation channels, we need to generalize the standard continuity equation such that the currents split into $ {m} \in \mathbb{N} $ distinct contributions. In case of the slit geometry the decay channels correspond to the longitudinal and transversal currents and we have, $ q_\mu^\alpha = q\delta_{\alpha \parallel} + Q_\mu \delta_{\alpha \perp} $. This splitting is physically motivated since we expect the relaxation dynamics to be significantly different in the directions parallel and orthogonal to the walls. From the above continuity equation we conclude that also the memory kernel naturally splits into multiple decay channels,
\begin{equation}\label{eq:contract}
K_{\mu \nu}(q,t) = \left[ \mathcal{C} \{\bm{\mathcal{K}} \}  \right]_{\mu \nu} \coloneqq \sum_{\alpha=1}^{m}\sum_{\beta=1}^{m} q_\mu^\alpha \mathcal{K}_{\mu \nu}^{\alpha \beta}(q,t) q_\nu^\beta,
\end{equation}
Here, the generalized memory kernel $\left[\bm{\mathcal{K}}(t)\right]^{\alpha\beta}_{\mu\nu}= \mathcal{K}_{\mu\nu}^{\alpha\beta}(q,t)$ carries in addition to the mode indices $\mu,\nu$ also the channel indices $\alpha,\beta \in \{1,\ldots,m\}$. Generally, we will denote matrix-valued objects with mode and channel indices by caligraphic letters.  The conventional current kernel $\mathbf{K}(t)$ is obtained as \emph{contraction}
$ \mathcal{C} \{\bm{\mathcal{K}}(t) \} $  from the generalized current kernel. 
We can perform a second Zwanzig-Mori projection step using the current modes $\{j_\mu^\alpha(\bm{q},t)\}$ as distinguished variable and derive evolution equations for the current kernels, 
\begin{equation}\label{eq:eom2}
\dot{\bm{\mathcal{K}}}(t) + \bm{\mathcal{J}}\bm{\mathcal{D}}^{-1}\bm{\mathcal{K}}(t)+ \int_0^t \bm{\mathcal{J}}\bm{\mathcal{M}}(t-t')\bm{\mathcal{K}}(t') \text{d}t'= 0,
\end{equation}
with the static current correlator $ {\mathcal{J}}^{\alpha \beta}_{\mu \nu}(q) \coloneqq N^{-1} \left\langle j^\alpha_\mu(\bm{q},0)^*j^\beta_\nu(\bm{q},0) \right\rangle ={\mathcal{K}}^{\alpha \beta}_{\mu \nu}(q,t=0) $. Here, the matrix $\bm{\mathcal{J}} \succ 0$ is positive definite, while the instantaneous damping  $ \bm{\mathcal{D}}^{-1} \succeq 0$ is positive semidefinite. The new memory kernel $\bm{\mathcal{M}}(t)$ formally corresponds to the correlation functions of generalized forces with the further reduced dynamic time evolution.  

The mode-coupling theory approach is now to approximate the force kernel $ \bm{\mathcal{M}}(t) $ as a bilinear functional of the intermediate scattering functions,
\begin{equation}\label{key}
\mathcal{M}_{\mu \nu}^{\alpha \beta}(q,t) = \mathcal{F}^{\alpha \beta}_{\mu \nu} \left[ \mathbf{S}(t),\mathbf{S}(t);q \right],
\end{equation}
with,
\begin{align}\label{eq:memory_mct}
&\hspace*{-0.4cm}\mathcal{F}^{\alpha \beta}_{\mu \nu} \left[ \mathbf{E},\mathbf{F};q \right] = \frac{1}{4N} \sum_{\bm{q}_1,\bm{q}_2=\bm{q}-\bm{q}_1} \sum_{\substack{\mu_1,\mu_2\\\nu_1,\nu_2}} \mathcal{Y}^\alpha_{\mu \mu_1\mu_2}(\bm{q},\bm{q}_1,\bm{q}_2) \\ &\hspace*{-0.9cm}\times(E_{\mu_1\nu_1}(q_1)F_{\mu_2\nu_2}(q_2)+F_{\mu_1\nu_1}(q_1)E_{\mu_2\nu_2}(q_2))\mathcal{Y}^\beta_{\nu \nu_1\nu_2}(\bm{q},\bm{q}_1,\bm{q}_2)^*,\nonumber
\end{align}
where the vertices $  \mathcal{Y}^\alpha_{\mu \mu_1\mu_2}(\bm{q},\bm{q}_1,\bm{q}_2) $ depend on the specific system that is described with this kind of MCT, like confined liquids \cite{Lang2012} or molecules \cite{Schilling:PRE_56_1997}. Here, we only need to assume that the vertices are smooth functions of the control parameters. We will anticipate the thermodynamic limit, 
however, for the moment we will assume that the wave vectors are discrete. Additionally, both the wave vectors and the modes will be truncated, i.e. the matrices are finite dimensional. 

To find a set of integro-differential equations similar to the ones used in previously performed asymptotic analysis for a single decay channel \cite{Franosch1997,Voigtmann2019} we introduce the Laplace transformation of 
a time-dependent  matrix $ \mathbf{A}(t) $,
\begin{equation}\label{key}
\text{LT} \left\{ \mathbf{A}(t) \right\}(z) = \hat{\mathbf{A}}(z) := {\rm i} \int_0^\infty \mathbf{A}(t) e^{{\rm i} zt} \text{d}t,
\end{equation}
and rewrite Eqs.~(\ref{eq:eom1}) and (\ref{eq:eom2}),
\newcommand{\disphat}[2][3mu]{\hat{#2\mkern#1}\mkern-#1}% \dispdot[<disp>]{<stuff>}
\begin{align}
	\hat{\mathbf{S}}(z) &= - [z\mathbf{S}^{-1}+\mathbf{S}^{-1}\hat{\mathbf{K}}(z)\mathbf{S}^{-1}]^{-1},\label{eq:eom1_Laplace}\\
\label{eq:eom2_Laplace}
	\hat{\bm{\mathcal{K}}}(z) &= - [z\bm{\mathcal{J}}^{-1}+{\rm i}\bm{\mathcal{D}}^{-1}+\disphat{\bm{\mathcal{M}}}(z)]^{-1}.
\end{align}
With this we can define an effective memory kernel, $ \mathbf{M}(z) $,  as,
\begin{equation}\label{eq:def_meff}
\hat{\mathbf{K}}(z) \eqqcolon -\left[ z\mathbf{J}^{-1} + {\rm i}\mathbf{{D}}^{-1}  + \hat{\mathbf{M}}(z)  \right]^{-1},
\end{equation}
with  static current correlator $ \mathbf{{J}}  = \mathbf{K}(t=0)= \mathcal{C}\{\bm{\mathcal{J}}\} \succ 0$ and effective damping matrix $ \mathbf{{D}}^{-1} =  \mathbf{J}^{-1}\mathcal{C}\{ \bm{\mathcal{J}}\bm{\mathcal{D}}^{-1}  \bm{\mathcal{J}}\} \mathbf{J}^{-1} \succeq 0  $.
One can show that $\hat{\mathbf{M}}(z)$ indeed shares all the properties of a matrix-valued correlation function \cite{Franosch_2014}, in particular, its spectrum is non-negative. Here, it corresponds precisely to the force kernel provided the second Zwanzig-Mori step is performed without splitting the currents. 

For the ISF in the time domain we therefore find the standard generalized harmonic oscillator equation for matrix-valued correlation functions,  
\begin{align}\label{eq:eom_eff}
\mathbf{J}^{-1}\ddot{\mathbf{S}}(t) &+\mathbf{{D}}^{-1} \dot{\mathbf{S}}(t)+\mathbf{S}^{-1}{\mathbf{S}}(t) + \int_{0}^{t} \mathbf{M}(t-t') \dot{\mathbf{S}}(t') \text{d}t' = 0, 
\end{align}
subject to the initial conditions $\mathbf{S}(t=0) = \mathbf{S}, \dot{\mathbf{S}}(t=0)=0$. 
The introduction of the effective memory kernel, $ \mathbf{M}(t) $, proved  also useful to derive a stable numerical integrator \cite{Chong2000,Gruber2019_2}. A similar scheme will be introduced for the schematic model in Sec.~\ref{sec:schematic}.

\subsection{\gj{Generalized evolution equations with parallel relaxation}}

\gj{The equations of motion (\ref{eq:eom1}), (\ref{eq:contract}), (\ref{eq:eom2}) and (\ref{eq:memory_mct}) form a closed set of equations that can describe systems beyond the example of simple liquids in confined geometry. Generally, the mode index $ \mu $ may refer to internal degrees of freedom (like orientations) or other broken symmetries. This includes systems like aspherical particles such as molecules (where the angular dependence of the microscopic densities is expanded in sphercial harmonics), active microrheology or active particles. In these systems parallel relaxation stems from different relaxation channels for transversal and rotational motion. Please note that although the mathematical structure of these equations, including matrix-valued correlation functions, is similar to previously published formalisms on mixtures  (corresponding to $m=1$), we consider identical particles in this manuscript (and focus on $m>1$). The formal similarity, however,  allows us to adapt proof strategies presented in Ref.~\citen{Voigtmann2019}.}

\subsection{Glass form factors}
\label{sec:glass_form}
The long-time limit of the correlation functions (for each wavenumber $q$)
\begin{equation}\label{key}
\mathbf{F} \coloneqq \lim\limits_{t\rightarrow \infty} \mathbf{S}(t),
\end{equation}
is referred to as glass form factor. Solutions with vanishing glass form factor are called ergodic or liquid, while non-vanishing glass form factors correspond to glass states. 
These long-time limits exist within the mode-coupling approximation under mild conditions even for Newtonian dynamics\cite{Franosch_2014} and rather obviously for overdamped dynamics\cite{Lang_2013}. In the case of matrix-valued correlation functions the glass form factors are necessarily positive-semidefinite matrices for each wavenumber~\cite{Franosch2002}.

The glass form factors $\mathbf{F}$ can be calculated without solving for the complete dynamics. The arguments can be literally transferred from the case of a single decay channel. 
 A non-vanishing glass form factor implies for the Laplace transform a pole $\hat{\mathbf{S}}(z) = \mathbf{F}/z + \{\text{smooth}\}$ and  the representation of the current kernel in terms of the effective memory kernel, Eq.~\eqref{eq:def_meff}, shows that 
\begin{align}\label{eq:nonergodic1}
 \mathbf{S} - \mathbf{F}  =  [ \mathbf{S}^{-1} + \mathbf{N} ]^{-1}
 ,
\end{align}
where $\mathbf{N} = - \lim_{z\to 0} z \hat{\mathbf{M}}(z) = \mathbf{M}(t\to\infty)$ corresponds to the long-time limit of the effective memory kernel. The abbreviation $\tilde{\mathbf{S}} := \mathbf{S}-\mathbf{F}\succ 0$ for the left hand side of Eq.\eqref{eq:nonergodic1} turns out to be useful for the further discussion. 
Assuming that all memory kernels reach a positive-definite long-time limit~\cite{Franosch2002,Lang2012} $ \mathcal{F}[\mathbf{F},\mathbf{F}] \succ 0 $, performing the contraction of the current kernel reveals that 
\begin{align}\label{eq:nonergodic2}
\mathbf{N} = \mathbf{N}[\mathbf{F}] = \left[ \mathcal{C}\{ \bm{\mathcal{{F}}}[\mathbf{F},\mathbf{F}]^{-1} \}  \right]^{-1}. 
\end{align}
More generally $\mathbf{N} = \mathbf{N}[\mathbf{E},\mathbf{F}]$ can be viewed as a functional with two distinct entries in the mode-coupling functional $\bm{\mathcal{F}}[\mathbf{E},\mathbf{F}]$. Then it has been shown in Ref.~[\citen{Lang2012}] that $\mathbf{N}[\mathbf{E},\mathbf{F}]$ maps positive-semidefinite matrices $\mathbf{E} \succeq 0, \mathbf{F} \succeq 0$ (for each $q$) to positive-semidefinite ones $\mathbf{N} \succeq 0$. As a consequence the long-time limit can be obtained by a convergent iteration scheme of Eqs.~\eqref{eq:nonergodic1}, \eqref{eq:nonergodic2} starting with $\mathbf{F}^{(0)} = \mathbf{S}$. The solution satisfies also the maximum principle and corresponds to the long-time limit of the MCT equations.

Upon changing the control parameters of the mode-coupling functional, the glass form factors will also change. The definition of the glass transition singularity implies that these changes are not smooth for smooth changes in the control parameters. To discuss the stability of the solutions in Eqs.~\eqref{eq:nonergodic1},\eqref{eq:nonergodic2}   the  functional $\tilde{\mathbf{S}}  \mathbf{N}[\mathbf{F}] \tilde{\mathbf{S}} $ 
is linearized at the fixed point solution, i.e the linear map $\mathbf{C}$ is introduced such that  $\tilde{\mathbf{S}}( \mathbf{N}[\mathbf{F}+\delta\mathbf{F}] - \mathbf{N}[\mathbf{F}]) \tilde{\mathbf{S}} = \mathbf{C}[\delta \mathbf{F}]+ {\cal O}(\delta \mathbf{F})^2$. From the explicit representation Eq.~\eqref{eq:nonergodic2} one finds     
\begin{align}\label{eq:Mlinearized}
\mathbf{C}[\delta\mathbf{F}] = 2  \tilde{\mathbf{S}}\mathbf{N}[\mathbf{F}]   \mathcal{C} \big \{   \bm{\mathcal{{F}}}[\mathbf{F},\mathbf{F}]^{-1} \bm{\mathcal{{F}}}[\delta \mathbf{F},\mathbf{F}] \bm{\mathcal{{F}}}[\mathbf{F},\mathbf{F}]^{-1}  \big\}   \mathbf{N}[\mathbf{F}] \tilde{\mathbf{S}} .
\end{align}
The linearization of Eq.~\eqref{eq:nonergodic1} then yields the condition
\begin{align}\label{eq:nonergodic_linearized}
 \delta \mathbf{F} = \mathbf{C}[\delta \mathbf{F}] + \Delta \mathbf{N},
\end{align}
where $\Delta \mathbf{N}[\mathbf{F}]$ is the small change of the functional $\mathbf{N}[\mathbf{F}]$  due to changes of control parameters while evaluated at the fixed point solution $\mathbf{F}$. Correspondingly, $\mathbf{C}[\delta \mathbf{F}]$ is  the proper generalization of the  stability matrix discussed in MCT for scalars \cite{Franosch2002,Gotze2009}.

Since the functional $\mathbf{N}[\mathbf{F}]$ for the multi-channel relaxation shares all the mathematical properties discussed in \cite{Franosch2002}  also the conclusions on the positive  linear map $\mathbf{C}[\delta \mathbf{F}]$ as inferred by the Frobenius-Perron theorem remain valid. In particular, it displays a maximal non-degenerate eigenvalue   $\gj{e}\leq 1$ with a positive-definite eigenvector. For eigenvalues $e<1$ the solution of Eq.~\eqref{eq:nonergodic_linearized} depends smoothly on control parameters $\delta \mathbf{F} \propto \Delta \mathbf{N}[\mathbf{F}]$, while $e=1$ corresponds to the glass transition singularity. 
The corresponding glass form factors at the singularity are denoted by $\mathbf{F}_{\text{c}} \succ 0$.
Furthermore the  eigenvector corresponding to such a critical point is denoted by $\mathbf{\tilde{H}} \succeq 0$, and  we indicate the fact that the control parameters are evaluated at the glass transition singularity by a subscript $\text{c}$, hence 
\begin{equation}\label{eq:cfp}
\mathbf{C}_\text{c}[\mathbf{\tilde{H}}] = \mathbf{\tilde{H}}. 
\end{equation}
For the moment this eigenvector is defined up to positive multiples. 

 As mentioned above, in deriving the properties of the effective memory kernel, we heavily relied on the assumptions that all memory kernels reach a positive-definite long-time limit $\mathcal{F}[\mathbf{F},\mathbf{F}] \succ 0$ such that the inverse can be performed in Eq.~\eqref{eq:nonergodic2}. 
 It is, however, conceivable  that parts of the parallel relaxation structure display type-A phenomenology (i.e. a continuous increase of the long-time limit as the transition is reached) \cite{Gruber2019_2} with vanishing glass form factor or that some of the relaxation channels remain ergodic \cite{Sentjabrskaja2016}. These cases will have to be studied separately.

\subsection{Equations of structural relaxation}

For the asymptotic analysis we assume that there exists a time window on a time scale $ t_\sigma \gg t_0 $ where the correlator $ \mathbf{S}(t) $ is close to $ \mathbf{F}_\text{c} $. Here, $ t_0 $ is an \emph{a priori} unknown time scale describing the initial transient dynamics. 
In this regime close to kinetic arrest, also the memory kernel $\bm{\mathcal{M}}(t)$ will remain almost constant, implying that $\hat{\bm{\mathcal{M}}}(z)$ becomes large, dominating the term $z \bm{\mathcal{J}} + {\rm i} \bm{\mathcal{D}}^{-1}$. 
We therefore can approximate 
\begin{equation} \label{eq:eom_K}
\hat{\mathbf{K}}(z) \approx -  \mathcal{C} \left\{ \disphat{\bm{\mathcal{M}}}(z)^{-1} \right\} = - \mathcal{C} \Big\{ \text{LT}  \left[ \bm{\mathcal{F}} \left[ \mathbf{S}(t),\mathbf{S}(t)\right] \right](z) \Big\}^{-1}
\end{equation}
For a single decay channel, the contraction is not needed   and we recover the equations considered for the analysis in Ref.~[\citen{Voigtmann2019}] for matrix-valued correlation functions.

\subsection{Asymptotic expansion}\label{Sec:Asymptotic_Expansion}

The strategy in this subsection is to expand  the channel-resolved memory kernel $\disphat{\bm{\mathcal{M}}}(z)$ in the window of the $\beta$-relaxation and adapt to the steps in Ref.~[\citen{Voigtmann2019}].  

First, we will expand the intermediate scattering function $ \mathbf{S}(t) $ close to the critical plateau $ \mathbf{F}_\text{c} $ on the divergent  time scale $ t_\sigma $. The precise definition of $t_\sigma$ 
will be elaborated only at the end of this subsection. 
For rescaled times $ \hat{t} = t/t_\sigma $ we thus identify $ \mathbf{S}(t) - \mathbf{F}_\text{c} $ as a small parameter which enables an asymptotic expansion.  In the following, it is assumed that the underlying bifurcation scenario is of type $ A_2 $, i.e. a  Whitney fold bifurcation, which is also the scenario discussed in the previous literature\footnote{For detailed discussions of the bifurcation scenario and its consequences we refer the reader to Ref.~\citen{Franosch1997} ($ l=2 $) and Ref.~\citen{Gotze2002} ($ l\geq 3 $)}. This leads to the ansatz for the asymptotic expansion,
\begin{equation}\label{eq:ansatz_expansion2}
{\mathbf{S}}({t}) = \mathbf{F}_\text{c} + \sqrt{\left| \sigma \right|} \mathbf{G}^{(1)}(t) + \sum_{n=2}^{\infty} \left|\sigma\right|^{n/2}\mathbf{G}^{(n)}(t),
\end{equation}
for some \emph{a priori} unknown separation parameter $ \sigma $. 
Additionally, we  assume a regular variation of the control parameters,
\begin{equation}\label{eq:ansatz_expansion3}
\mathbf{S} = \mathbf{S}_\text{c} + \sigma \mathbf{S}^{(1)} + \mathcal{O}(\sigma^2),
\end{equation}
which applies similarly to the mode-coupling functional~$ \bm{\mathcal{F}} $. The procedure is now to expand $\disphat{\bm{\mathcal{M}}}(z)$ in powers of $\sqrt{|\sigma|}$, and then by Eq.~(\ref{eq:eom_K}) also the current kernel $\hat{\mathbf{K}}(z)$, and finally with Eq.~(\ref{eq:eom1_Laplace}) the density correlation function $\hat{\mathbf{S}}(z)$.   
For details to the following derivation we refer the reader to Refs.~[\citen{Franosch1997,Voigtmann2019}] since we will mainly focus on the discussion of differences to previously published results.

\subsubsection{Zeroth order expansion}

\noindent To zeroth order in $ \sqrt{\left| \sigma \right|} $ we find an equation for the critical nonergodicity parameter, 
\begin{equation}\label{key}
\mathbf{F}_\text{c}  = \left[ \mathbf{S}_\text{c}^{-1} +  \mathbf{S}_\text{c}^{-1} \mathbf{N}_\text{c}^{-1} \mathbf{S}_\text{c}^{-1} \right]^{-1}
=\mathbf{S}_\text{c} - \left[ \mathbf{S}^{-1}_\text{c} + \mathbf{N}_\text{c} \right]^{-1}, 
\end{equation}
which is equivalent to the expression derived in Ref.~[\citen{Lang2010}] Eq.~(11). The expression for $ \mathbf{N}_\text{c} $ is given by,
\begin{equation}\label{eq:Fc}
\mathbf{N}_\text{c}^{-1}=\mathcal{C}\{ \bm{\mathcal{{F}}}_\text{c}[\mathbf{F}_\text{c},\mathbf{F}_\text{c}]^{-1} \} .\\
\end{equation}
As shown in Sec.~\ref{sec:glass_form} the two equations above readily yield an iteration scheme that is guaranteed to converge to the long-time limit of the dynamic equations and thus to the physically desired solution (see also Refs.~[\citen{Lang2012},\citen{Lang_2013}]). 

\subsubsection{First order expansion}

\noindent To first order we obtain (details of the expansion are  presented in App.~\ref{app:exp}),
\begin{equation}\label{eq:eigenvector}
\mathbf{G}^{(1)}(t) - \mathbf{C}_\text{c}[\mathbf{G}^{(1)}(t)]=0,
\end{equation}
with the linear map,
\begin{equation}\label{key}
  \mathbf{C}_\text{c}[\mathbf{G}^{(1)}(t)] \coloneqq 2\tilde{\mathbf{S}}_c \mathbf{N}_\text{c}^{(1)}[\mathbf{G}^{(1)}(t),\mathbf{F}_\text{c}]  \tilde{\mathbf{S}}_c,
\end{equation}
and,
\begin{align}\label{eq:FcG}
\mathbf{N}^{(1)}_\text{c}[\mathbf{E},\mathbf{F}] = \mathbf{N}_\text{c}   \mathcal{C} \big \{   \bm{\mathcal{{F}}}_\text{c}[\mathbf{F}_\text{c},\mathbf{F}_\text{c}]^{-1} \bm{\mathcal{{F}}}_\text{c}[\mathbf{E},\mathbf{F}] \bm{\mathcal{{F}}}_\text{c}[\mathbf{F}_\text{c},\mathbf{F}_\text{c}]^{-1}  \big\}   \mathbf{N}_\text{c}.
\end{align}
Comparing Eq.~\eqref{eq:eigenvector} with Eq.~\eqref{eq:cfp} shows immediately that $ \mathbf{G}^{(1)}(q,t) $ is proportional to the critical Perron-Frobenius eigenvector\cite{Franosch1997,Gotze2009} $ \tilde{\mathbf{H}} $ for all times  in the window of the $\beta$-relaxation. The time dependence thus separates from the wave vector and mode index dependence and we find, 
\begin{equation}\label{eq:factorization}
\mathbf{G}^{(1)}(q,t) = \tilde{\mathbf{H}}(q) \tilde{g}(\hat{t}), 
\end{equation}
where we reinstated the dependence on rescaled times  $\hat{t} = t/t_\sigma$ in the $\beta$-relaxation.
Eq.~(\ref{eq:factorization}) is also called the factorization theorem, including the critical amplitude $ \tilde{\mathbf{H}}(q) $ and the $\beta$-correlator $ \tilde{g}(\hat{t}) $. It basically states that close to the critical point on the time scale of $ t_\sigma $ all dynamical correlation functions can be rescaled such that they superimpose on a single universal master curve. In the following, we will fix this eigenvector uniquely using the normalizations,
\begin{align}\label{eq:norm}
\text{tr}\left( \hat{\mathbf{H}} \tilde{\mathbf{H}} \right) &= 1,\nonumber\\
\text{tr}\left( \hat{\mathbf{H}} \tilde{\mathbf{H}} \tilde{\mathbf{S}}_\text{c}^{-1}  \tilde{\mathbf{H}} \right) &= 1.
\end{align}
The multiplication of the matrices is to be understood separately for each wavenumber, while the trace is extended to account for the set of wavenumbers $\text{tr}\left(\mathbf{A} \right) = \sum_q \text{tr}  \mathbf{A}(q) $. We also introduced $ \hat{\mathbf{H}} $ as the corresponding left-eigenvector of the above discussed linear map, defined as $ \text{tr}( \hat{\mathbf{H}} \mathbf{f} ) = \text{tr}( \hat{\mathbf{H}} \mathbf{C}[\mathbf{f}] )$ for any trial matrix $ \mathbf{f} $.\cite{Voigtmann2019}

\subsubsection{Second order expansion}

\noindent To second order we find,
\begin{align}\label{eq:2nd_order}
\mathbf{G}^{(2)}({t}) &- \mathbf{C}_\text{c}[\mathbf{G}^{(2)}({t})] = \tilde{\mathbf{S}}_\text{c} \mathbf{N}_\text{c}^{(1)}[\tilde{\mathbf{H}},\tilde{\mathbf{H}}]  \tilde{\mathbf{S}}_\text{c}g(\hat{t})^2  - 2\tilde{\mathbf{S}}_\text{c}\mathbf{N}_\text{c}^{(1)}[\tilde{\mathbf{H}},\mathbf{F}_\text{c}] \tilde{\mathbf{H}} \frac{\text{d}}{\text{d} \hat{t}} (g \ast g) (\hat{t})  \nonumber\\
&+ \tilde{\mathbf{S}}_\text{c} \mathbf{A}  \tilde{\mathbf{S}}_\text{c} \frac{\text{d}}{\text{d} \hat{t}} (g \ast g) (\hat{t}) 
+\tilde{\mathbf{S}}_\text{c}\mathbf{S}_\text{c}^{-1} \left(  \mathbf{S} \mathbf{N}_\epsilon(\mathbf{S}-\mathbf{F}_\text{c}) -\mathbf{S}_\text{c} \mathbf{N}_\text{c}\tilde{\mathbf{S}}_\text{c}   \right)/|\sigma|,
\end{align}
with,
\begin{equation}\label{eq:def_A}
\mathbf{A} = 4 \left (  \mathbf{N}^{(1)}_\text{c}[\tilde{\mathbf{H}},\mathbf{F}_\text{c}] \mathbf{N}_\text{c}^{-1}  \mathbf{N}^{(1)}_\text{c}[\tilde{\mathbf{H}},\mathbf{F}_\text{c}] -\mathbf{N}_\text{c}^{(2)}[\tilde{\mathbf{H}},\tilde{\mathbf{H}}] \right),
\end{equation}
and the definitions,
\begin{align}
\mathbf{N}_\epsilon&=\left[ \mathcal{C} \{  \bm{\mathcal{F}}[\mathbf{F}_\text{c},\mathbf{F}_\text{c}]^{-1}  \}\right]^{-1}, \label{eq:Mnoncrit}\\
\mathbf{N}_\text{c}^{(2)}[\mathbf{E},\mathbf{F}] &= \mathbf{N}_\text{c}  \mathcal{C} \big\{ \bm{\mathcal{{F}}}_\text{c}[\mathbf{F}_\text{c},\mathbf{F}_\text{c}]^{-1} \bm{\mathcal{{F}}}_\text{c}[\mathbf{E},\mathbf{F}_\text{c}] \bm{\mathcal{{F}}}_\text{c}[\mathbf{F}_\text{c},\mathbf{F}_\text{c}]^{-1}
\bm{\mathcal{{F}}}_\text{c}[\mathbf{F},\mathbf{F}_\text{c}] \bm{\mathcal{{F}}}_\text{c}[\mathbf{F}_\text{c},\mathbf{F}_\text{c}]^{-1}  \big\} \mathbf{N}_\text{c}.\label{eq:M2}
\end{align}
Details of the expansion are presented in App.~\ref{app:exp} and the factorization theorem (\ref{eq:factorization}) has been used. One readily checks that the matrix $ \mathbf{A} $ evaluates to  zero if there is only a single relaxation channel. The last term in Eq.~(\ref{eq:2nd_order}) collects the linear terms in $ \left|\sigma \right|$ due to variations of the control parameters. We now follow the standard route utilizing the solubility conditions \cite{Franosch1997,Voigtmann2019}, 
\begin{equation}\label{key}
\text{tr}\big( \hat{\mathbf{H}} \mathbf{G}^{(n)}({t}) - \hat{\mathbf{H}} \mathbf{C}_\text{c}[\mathbf{G}^{(n)}({t})] \big)= 0 = \text{tr}\big( \hat{\mathbf{H}} \mathbf{I}^{(n)} \big),
\end{equation}
and identifying the different terms in Eq.~(\ref{eq:2nd_order}) as,
\begin{align}\label{key}
\sigma &= \text{tr} \left[ \hat{\mathbf{H}}\tilde{\mathbf{S}}_\text{c}\mathbf{S}_\text{c}^{-1} \left(  \mathbf{S} \mathbf{M}_\epsilon(\mathbf{S}-\mathbf{F}_\text{c}) -\mathbf{S}_\text{c} \mathbf{M}_{c}\tilde{\mathbf{S}}_\text{c}  \right)  \right],\\
\tilde{\lambda} &= \text{tr} \left[ \hat{\mathbf{H}}  \tilde{\mathbf{S}}_\text{c} \mathbf{N}_\text{c}^{(1)}[\tilde{\mathbf{H}},\tilde{\mathbf{H}}] \tilde{\mathbf{S}}_\text{c} \right],\\
\Delta &= \text{tr} \left[ \hat{\mathbf{H}} \tilde{\mathbf{S}}_\text{c} \mathbf{A}  \tilde{\mathbf{S}}_\text{c} \right], \label{eq:Delta_term}
\end{align}
to find,
\begin{align}\label{eq:beta_scaling}
\left(1- \Delta \right) \frac{\text{d}}{\text{d} \hat{t}} (\tilde{g} \ast \tilde{g}) (\hat{t}) =\tilde{\lambda} \tilde{g}(\hat{t})^2 + \text{sgn}\, \sigma,
\end{align}
with the convolution,
\begin{equation}\label{key}
(f \ast g)(t) = \int_0^t f(t-t^\prime)g(t^\prime)\text{d}t^\prime.
\end{equation}
In case of a single decay channel, the new term does not show up, $ \Delta=0 $, and the result reduces to the well-known $ \beta $-scaling equation derived in Refs.~[\citen{Gotze2002,Voigtmann2019,Franosch1997}]. To recover their result in the presence of multiple relaxation channels we employ \emph{a posteriori} a rescaling,
\begin{align}\label{key}
{g}(\hat{t}) =  \tilde{g}(\hat{t})\sqrt{1-\Delta},\\
{\mathbf{H}}(q)=  \tilde{{\mathbf{H}}}(q)/\sqrt{1-\Delta},\\
{\lambda} = \tilde{\lambda}/(1-\Delta).
\end{align}
This is possible due to the scale invariance of the scaling equation and it also leaves the factorization theorem untouched. This eventually leads to the original $ \beta $-scaling equation,
\begin{align}\label{eq:beta_scaling2}
 \frac{\text{d}}{\text{d} \hat{t}} ({g} \ast {g}) (\hat{t}) ={\lambda} {g}(\hat{t})^2 + \text{sgn}\, \sigma.  
\end{align}
The scaling equation displays a  power-law solution with exponents $1/2 \leq -a \leq  0$ for short rescaled times $\hat{t} \ll 1$. For $\text{sgn } \sigma = -1$ a second power law with exponent $0 \leq b \leq 1$ emerges at long rescaled times $\hat{t}\gg 1$,  while the scaling function converges for long times to a finite value for $\text{sgn } \sigma = 1$. The exponent parameter 
 $ \lambda $, connects the exponents, $ a,b $, of the $ \alpha $- and $ \beta $-relaxation processes via \emph{G\"otze's exponent relation},
\begin{equation}\label{key}
\frac{\Gamma(1+b)^2}{\Gamma(1+2b)} = \lambda = \frac{\Gamma(1-a)^2}{\Gamma(1-2a)}.
\end{equation}
To fix a unique solution of the scaling equation, the condition  $g(\hat{t} \ll 1) = \hat{t}^{-a}$ is imposed.  

\subsubsection{Critical dynamics}
\label{ch:scaling}

We recall the most important conclusions for the critical dynamics close to the glass transition that can be drawn from Eqs.~(\ref{eq:ansatz_expansion2}),(\ref{eq:factorization}) and (\ref{eq:beta_scaling2}) (see Refs.~[\citen{Franosch1997,Gotze2009}] for extended derivations). 

\begin{itemize}
	\item For $ t \gg t_0 $ and $ {t} \ll t_\sigma $ the short-time solution $ g(\hat{t} \ll 1) = \hat{t}^{-a} $ sets $ t_\sigma = t_0 \left| \sigma\right|^{-1/2a} $ and we find,
	\begin{align}\label{eq:crit_asymptote}
	\mathbf{S}(q,t) &= \mathbf{F}_\text{c}(q) + \mathbf{{H}}(q) (t/t_\sigma)^{-a} \sqrt{\left|\sigma \right|} + \mathcal{O}(\sqrt{\left|\sigma \right|}^2),\\
	\boldsymbol{\chi}''(q,\omega) &= \mathbf{{H}}(q)  \Gamma(1-a)\sin\left(\pi a/2\right)(\omega t_\sigma)^a\sqrt{\left|\sigma \right|}+ \mathcal{O}(\sqrt{\left|\sigma \right|}^2) .\label{eq:crit_asymptote_freq}
	\end{align}
	Here, $ \boldsymbol{\chi}''(q,\omega) = \omega \mathbf{S}''(q,\omega) $ is the susceptibility spectrum, calculated from the correlation spectrum, $ \mathbf{S}''(q,\omega) = \int_0^\infty \cos (\omega t ) \mathbf{S}(q,t) \text{d}t. $
	
	\item For $ \sigma \geq 0 $ and $ {t} \gg t_\sigma $ the asymptotic expansion of the glass form factor follows,
	\begin{equation}\label{eq:static_asymptote}
	\mathbf{F}(q) = \lim\limits_{t\rightarrow \infty }	\mathbf{S}(q,t) = \mathbf{F}_\text{c}(q) + \mathbf{{H}}(q) \sqrt{\frac{\sigma}{1-\lambda}}+ \mathcal{O}(\sqrt{\left|\sigma \right|}^2).	
	\end{equation}
	\item For $ \sigma < 0 $ and $  {t} \gg t_\sigma $ we find the emergence of a second power law, $ g(\hat{t} \gg 1 ) = -B\hat{t}^b $, which sets the $ \alpha $-relaxation time $ t'_\sigma = (t_0/B^{1/b})\left| \sigma \right|^{-\gamma} $, with $ \gamma = 1/2a + 1/2b $. For the intermediate scattering function and the susceptibility spectrum we obtain,
	\begin{align}\label{eq:alpha_asymptote}
	\mathbf{S}(q,t) &= \mathbf{F}_\text{c}(q) - \mathbf{{H}}(q) (t/t'_\sigma)^{b}+ \mathcal{O}(\sqrt{\left|\sigma \right|}^2),\\
	\boldsymbol{\chi}''(q,\omega) &= \mathbf{{H}}(q)  \Gamma(1+b)\sin\left(\pi b/2\right)(\omega t'_\sigma)^{-b}+ \mathcal{O}(\sqrt{\left|\sigma \right|}^2).\label{eq:alpha_asymptote_freq}
	\end{align}
\end{itemize}

\subsection{Discussion}
\label{sec:discussion}

We find that the existence of multiple decay channels has no fundamental influence on the asymptotic scaling laws which can be derived from various kinds of mode-coupling theories.  This confirms that the standard scaling analysis can indeed also be applied to systems where multiple decay channels emerge naturally, such as molecules, confined fluids or active particles. This result has been anticipated before \cite{Winkler2000} but to the best knowledge of the authors it has not yet been {shown explicitly}. 

The microscopic expression of the exponent parameter differs from the case of the single relaxation channel due to a new contribution as highlighted in the term $\Delta$, Eq.~\eqref{eq:Delta_term}.  
These non-trivial terms could also be hidden in the normalizations of the eigenvectors Eq.~(\ref{eq:norm}) at the prize of lengthy expressions. In both cases the derivation shows that the difference introduced in Eq.~(\ref{eq:def_A}), which is largest if the decay channels are very different, will have an influence on the emergent power-law exponents.

\section{Schematic model with two decay channels}
\label{sec:schematic}

To understand the emergent static and dynamic properties of mode-coupling theory, schematic models have been proved to be very useful \cite{Krieger1987,Gotze2009}. These models are characterized by a very small number of modes $ M $ (usually $ M \leq 2 $) which significantly simplifies and accelerates the numerical solution of the equations. Inspired by the MCT equations for liquids in confinement, we suggest a schematic model to study the impact of multiple decay channels. We restrict ourselves to diagonal $ 2\times 2 $ matrices with two distinct decay channels, leading to the equations of motion,
 \begin{align}
\hat{S}_i(z) &= -\left[z+\hat{K}_i(z)\right]^{-1}, \qquad \vspace*{3cm} i = 1,2,\label{eq:mod1}\\
\hat{K}_i(z) &= Q^\parallel \hat{\mathcal{K}}^\parallel_i(z) + Q^\perp \hat{\mathcal{K}}^\perp_i(z),\label{eq:mod2}\\
\hat{\mathcal{K}}_i^\alpha(z)&=-\left[ z + {\rm i} \nu_i+  \hat{\mathcal{M}}^\alpha_i(z)\right]^{-1},  \quad \,\,\alpha = \parallel,\perp.\label{eq:mod3}
\end{align}
Here, Eqs.~(\ref{eq:mod1}),(\ref{eq:mod2}) and (\ref{eq:mod3}), are the direct equivalents of Eqs.~(\ref{eq:eom1_Laplace}),(\ref{eq:contract}) and (\ref{eq:eom2_Laplace}), respectively, where $ i \equalhat (q,\mu) $. For the memory kernels, we now use the Bosse-Krieger model\cite{Krieger1987},
 \begin{align}
\mathcal{M}_1^\alpha(t)&= Q^\alpha k\left[\delta^\alpha S_1(t)^2+(1-\delta^\alpha)S_2(t)^2\right] = Q^\alpha \tilde{\mathcal{M}}_1^\alpha(t)\label{eq:schematic_mem1},\\
\mathcal{M}_2^\alpha(t)&= Q^\alpha krS_1(t)S_2(t) = Q^\alpha \tilde{\mathcal{M}}_2^\alpha(t)\label{eq:schematic_mem2}.
\end{align}
Here, the parameters $ 0 \leq \delta^\parallel, \delta^\perp \leq 1 $ and $ 0 < r < \infty $ are assumed to be fixed and $ k $ is the control parameter. If one of the decay channels, $ Q^\parallel, Q^\perp, $ is zero, the schematic model reduces to the well-known single-channel Bosse-Krieger model \cite{Krieger1987}. In the following, we will use $ \delta^\parallel = 0.9 $ and $ \delta^\perp = 0.1 $. The two decay channels therefore differ such that in parallel ``direction'' mode 1 couples stronger to itself than to mode 2, while in perpendicular ``direction'' it couples weakly to itself. To increase the stability and avoid unrealistic oscillations we include an instantaneous  damping $ \nu_1 > 0  $ and $\nu_2 = 0$, as is commonly done in MCT \cite{Franosch1997,Gotze2009}.

Eqs.~(\ref{eq:mod1})-(\ref{eq:mod3}) together with the expressions for the memory kernels Eq.~(\ref{eq:schematic_mem1}) and Eq.~(\ref{eq:schematic_mem2}) can be solved numerically by combining standard techniques for bulk MCT \cite{Sperl2000} and methods that were recently suggested to solve MCT for active microrheology \cite{Gruber2016,Gruber2019_2}. Details can be found in Appendix \ref{ap:numerics}. In Table~\ref{tab:parameters} the parameters for the model used in this work are summarized.

\renewcommand{\arraystretch}{1.1}
\begin{table}
	\centering \begin{tabular}{cc|cc|cc} 
		$ \delta^\parallel $ & \hphantom{1111} $ 0.9 $ \hphantom{1111} & $ k^\text{c} $ & 42.7021257192(1) & $ \tilde{\lambda} $ &  \hphantom{11} 0.6871  \hphantom{11} \\
		\myrowcolour
		$ \delta^\perp $ & $ 0.1 $ & $ F_1^\text{c} $ & 0.7555& $ {\lambda} $ &  0.7039  \\
		$ r $ & $ 0.075 $ & $ F_2^\text{c} $ &   0.1734 & $a $ & 0.3322 \\
		\myrowcolour
		$ Q^\parallel $ & $ 1.0 $ & $ {H}_1 $ &  0.5177 & $ b $ & 0.6622\\ 
		$ Q^\perp $ & $ 1.0 $ & $ {H}_2 $ &  0.5664 &$ B $ & 0.6872   \\ 
		\myrowcolour
		$ \nu_1 $ & $ 4.0 $ &  $ \Delta   $ & -0.0244 & $ t_0 $ & 0.0371 \\
	\end{tabular} 
	\caption{Summary of the parameters and critical properties of the schematic Bosse-Krieger model with two decay channels as defined in Eqs.~(\ref{eq:mod1})-(\ref{eq:schematic_mem2}). The parameter $ B $ was determined by linear interpolation of Tab.~3 in Ref.~[\citen{Gotze_1990}]. }
	\label{tab:parameters}
\end{table} 
\renewcommand{\arraystretch}{1.0}

\subsection*{Statics: Nonergodicity parameter}

\begin{figure}
	\includegraphics[scale=1.0]{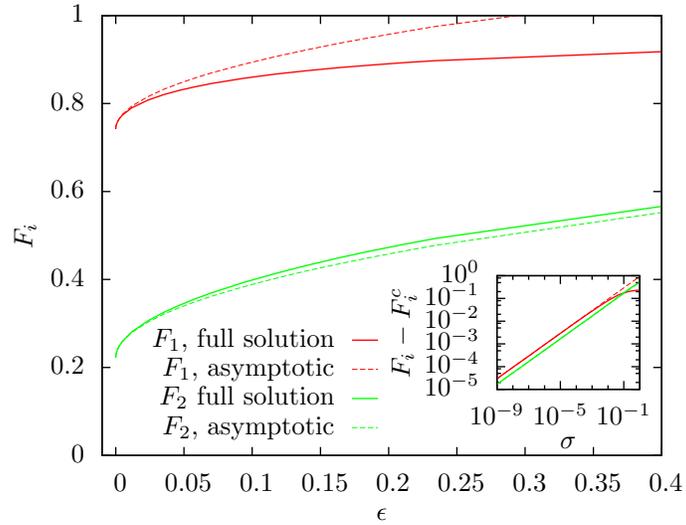}
	\caption{Nonergodicity parameters $ F_i $ for various values of the {control-parameter distance} $ \epsilon = (k-k^\text{c})/k^\text{c}. $ The asymptotes were determined according to Eq.~(\ref{eq:static_asymptote}). The inset shows the variation from the critical nonergodicity parameters in dependence of the separation parameter $ \sigma $ in double logarithmic representation. } 
	\label{fig:nonergo}
\end{figure}

The bifurcation of the nonergodicity parameter above the critical point is displayed in 
Fig.~\ref{fig:nonergo}. As anticipated, the curves can be described by the leading order asymptote, $ F_i \propto \sqrt{\epsilon} = \sqrt{ (k-k^\text{c})/k^\text{c}} $, for small control-parameter distance $\epsilon$ to the transition (see also Eq.~(\ref{eq:static_asymptote})). The inset provides numerical evidence that the expansion presented in the previous chapter indeed accurately describes the asymptotic behavior of a system with multiple decay channels. 

\subsection*{Dynamics: Correlators and susceptibility}

\begin{figure}
	\hspace*{-1.2cm}\includegraphics[scale=0.95]{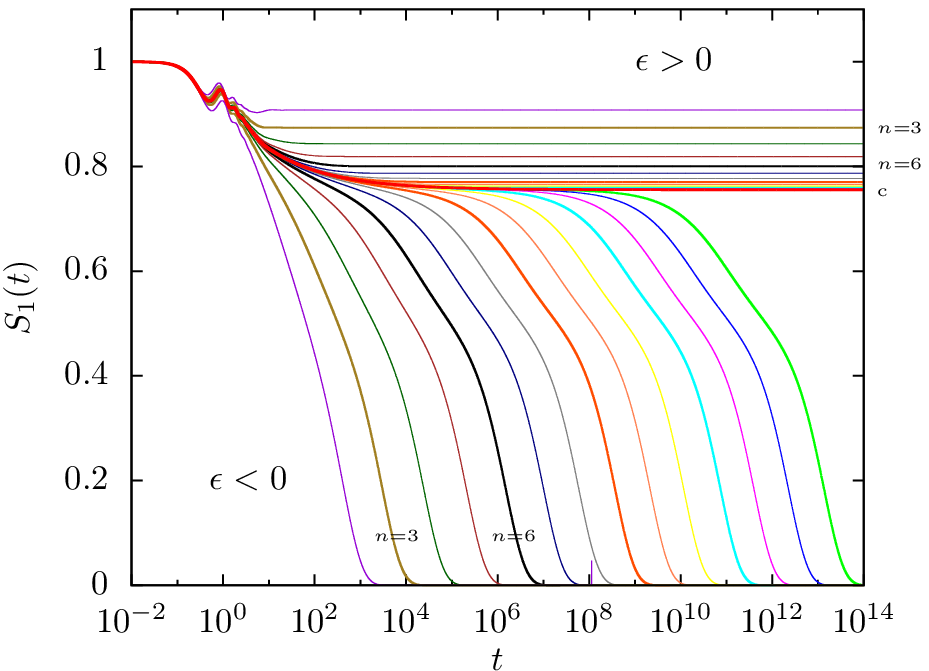}\includegraphics[scale=0.95]{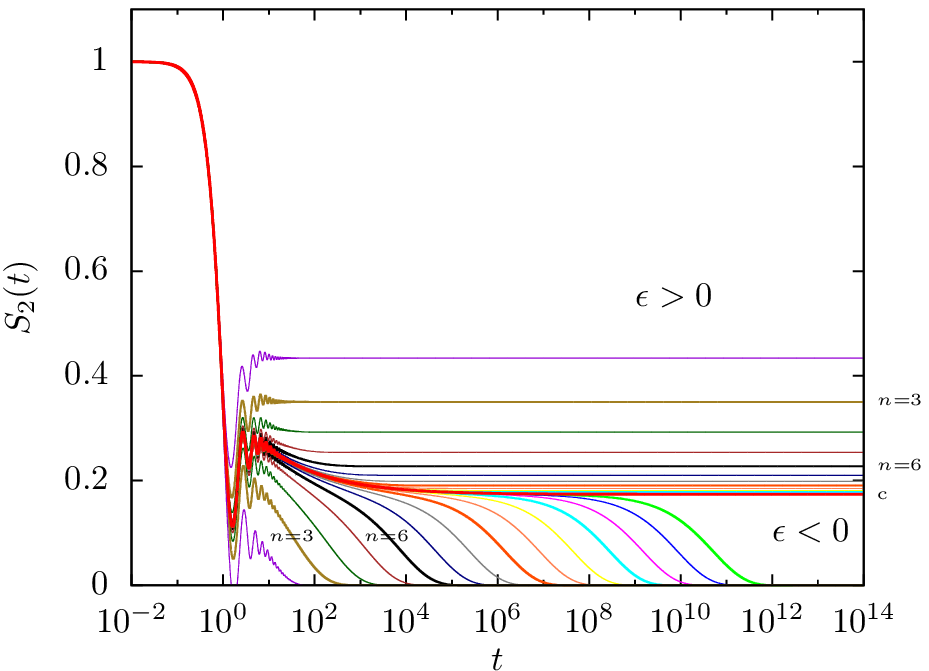}
	\caption{ Time-dependent correlators $ S_1(t) $ (left) and $ S_2(t) $ (right). The different curves are calculated for control parameters $ \epsilon=\pm 10^{-n/3} $ ($ n $ increases from left to right for $ \epsilon<0 $ and from top to bottom for $ \epsilon > 0 $ ). The critical correlator for $ k=k^\text{c} $ ($ \epsilon = 0 $) is displayed as a  thick line and labeled by ``c''. } 
	\label{fig:time_full}
\end{figure}

The time dependences of the correlators $ S_i(t) $ are visualized in Fig.~\ref{fig:time_full}. Both curves show qualitatively the same time dependence as known from bulk MCT, i.e. they are smooth functions of $ \epsilon $ on any finite interval of time with divergent time scales, $ t_\sigma $ and $ t_\sigma^\prime $, when approaching the critical point, $ \epsilon \rightarrow 0 $. For a detailed discussion of these general features we refer the reader to Ref.~[\citen{Franosch1997}]. More quantitatively it can be observed that the Markovian contribution to the memory kernel leads to a damping of the correlator $ S_1(t) $ on a time scale $ \nu_1^{-1} $, as expected from the equations of motion. This becomes even more apparent from the appearance of strong oscillations in the dynamics of correlator $ S_2(t) $  for which no Markovian damping was included.

\begin{figure}
	\hspace*{-1.2cm}\includegraphics[scale=0.95]{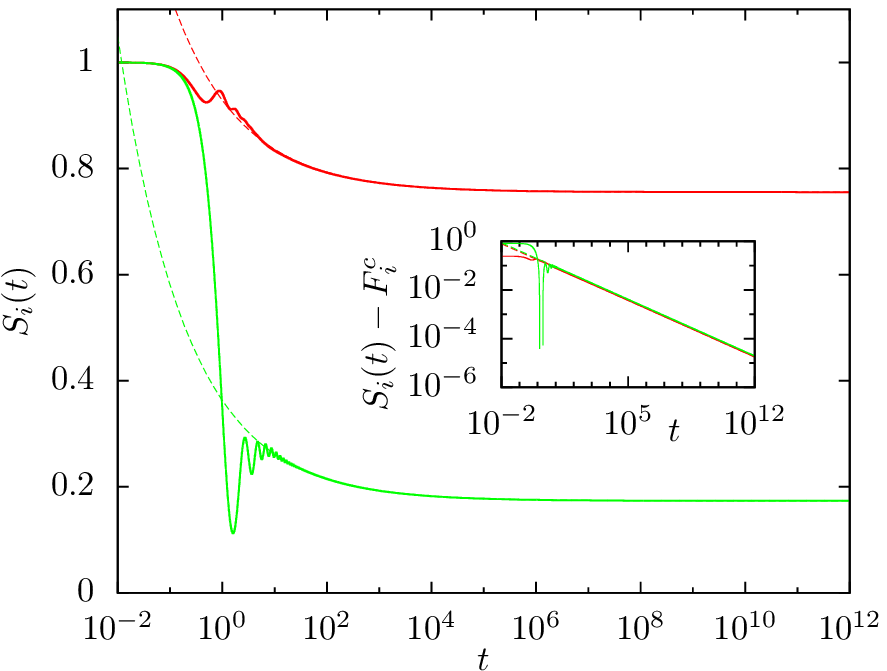}\includegraphics[scale=0.95]{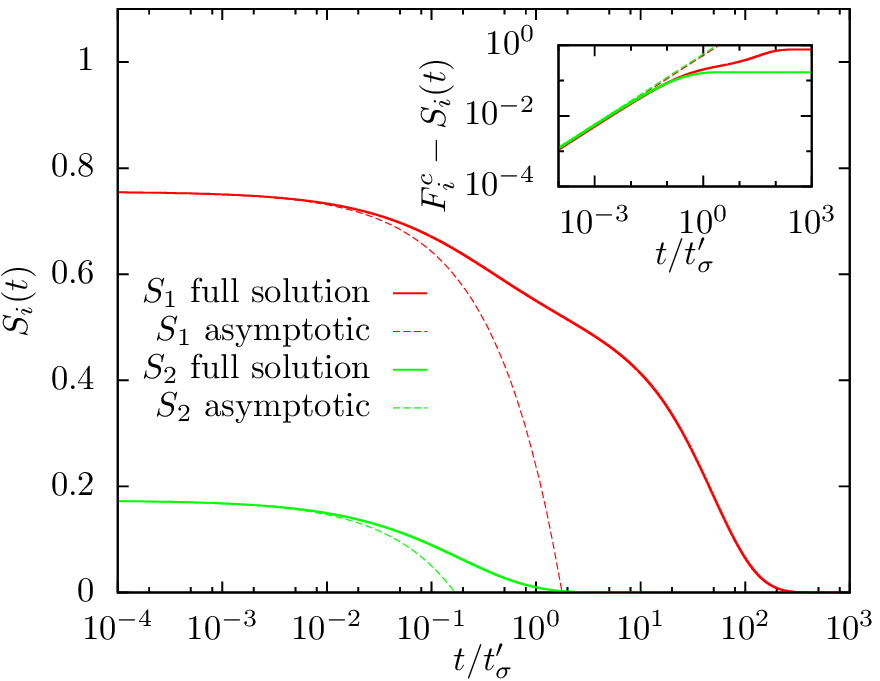}
	\caption{Time dependence of the critical correlators $ S_i(t) $ (left) and the $ \alpha $-relaxation process (right). The asymptotes in the left panel ($ \beta $-relaxation) were calculated according to Eq.~(\ref{eq:crit_asymptote}). The time scale $ t_0 $ was determined by matching the asymptotic solution of $ S_1(t) $. The inset shows the convergence to the critical nonergodicity parameter in double logarithmic representation. Similarly, the right panel shows the $ \alpha $-relaxation with the asymptotes calculated according to Eq.~(\ref{eq:alpha_asymptote}) for a separation parameter $ \sigma = 6.36 \cdot 10^{-9} $ and thus $ t_\sigma^\prime=2.182\cdot 10^{17} $.   } 
	\label{fig:time_crit}
\end{figure}

The $ \alpha$- and $ \beta $-relaxation processes are shown separately and in more detail in Fig.~\ref{fig:time_crit}, including the asymptotic scaling laws derived in Ch.~\ref{ch:scaling}. The $ \beta $-relaxation of the critical correlator is depicted in the left panel. For $ t > 10^2 $ there is no visible difference between the time fractal, $ g(\hat{t}) = \hat{t}^{-a}  $, as derived in Eq.~(\ref{eq:crit_asymptote}), and the correlator $ S_i(t,\epsilon \rightarrow +0 ) $. Here, we  emphasize  that the emergent power-law exponent $ a $ is a non-trivial result of the asymptotic expansion in a scenario with multiple decay channels presented in this manuscript. When approaching the critical point from below, a second relaxation process, called $ \alpha $-relaxation, is observed (see right panel of Fig.~\ref{fig:time_crit}). This decay from the plateau value $ F_i^\text{c} $ is described by a second power law, $ g(\hat{t}) = - B \hat{t}^{b}  $, which is commonly referred to as  von Schweidler law \cite{Gotze2009}. Interestingly, a distinct ``kink'' is observed in the $ \alpha $-relaxation at   $t/t_\sigma'\approx 10$ that has not yet been reported for MCT dynamics. We will postpone its discussion to the next paragraph.

\begin{figure}
	\hspace*{-1.2cm}\includegraphics[scale=0.95]{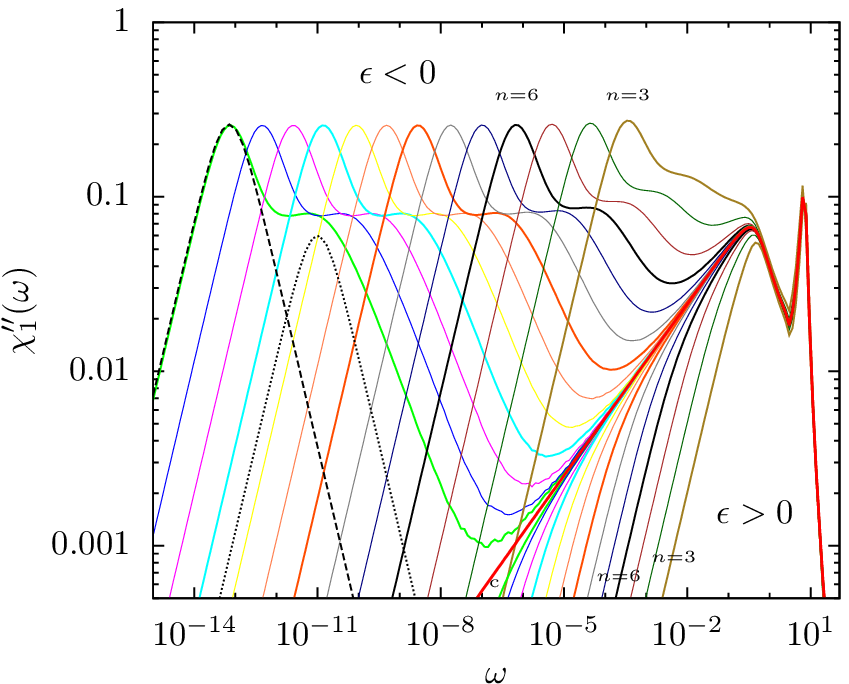}\includegraphics[scale=0.95]{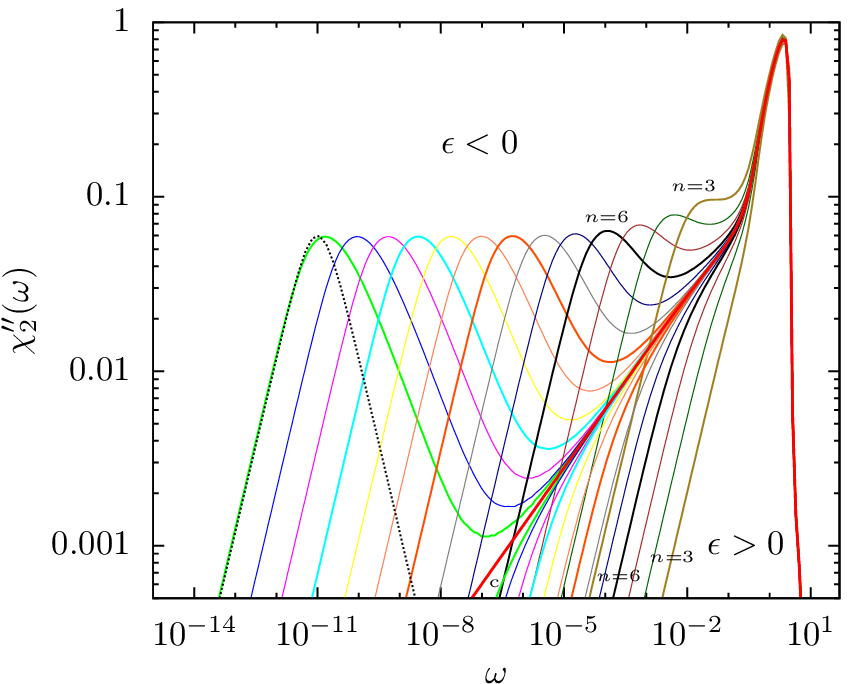}
	\caption{Frequency-dependent susceptibility $ \chi_1^{\prime \prime}(\omega) = \omega {S}_1^{\prime \prime}(\omega) $ (left) and $ \chi_2^{\prime \prime}(\omega) $ (right) ($ n $ increases from right to left)
	  for the same  parameters as in Fig.~\ref{fig:time_full}. The dashed and dotted, black lines show Debye peaks, $ \chi^{\prime \prime}_{\text{D}_i}(\omega) = 2 \chi_{\text{max}_i} \omega \tau_{ \text{\tiny D}_i}/ \left[ 1+ (\omega \tau_{ \text{\tiny D}_i} )^2  \right] $, for mode 1 and 2, respectively ($ \chi_{\text{max}_1} = 0.26 $, $ \tau_{ \text{\tiny D}_1} = 1.4\cdot 10^{13} $, $ \chi_{\text{max}_2} = 0.06 $, $ \tau_{ \text{\tiny D}_2}= 1.0\cdot 10^{11} $). The dotted line is also included to the spectrum $ \chi_1^{\prime \prime}(\omega)  $ for the sake of visualization. } 
	\label{fig:susc_full}
\end{figure}

\begin{figure}
	\hspace*{-1.2cm}\includegraphics[scale=0.95]{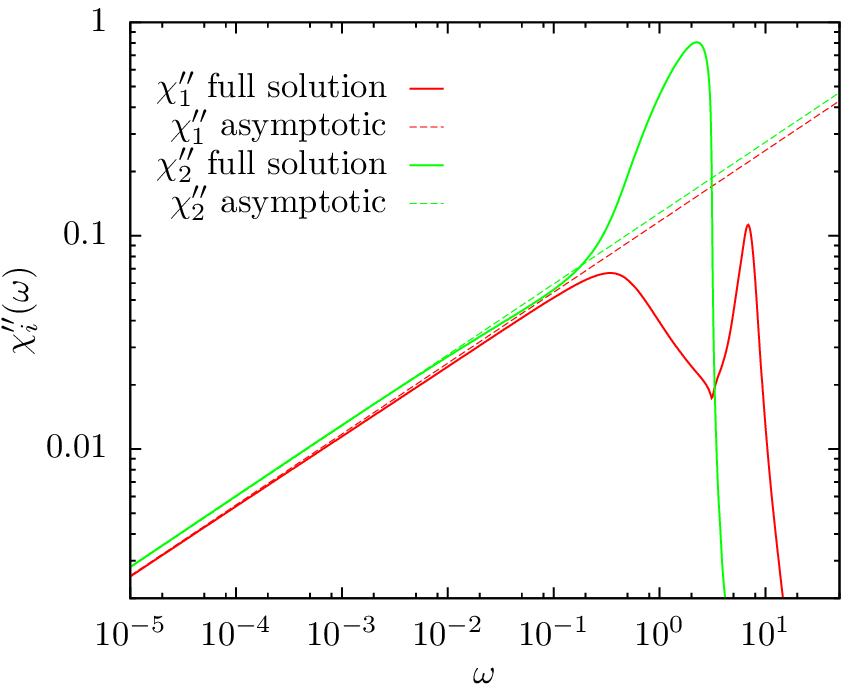}\includegraphics[scale=0.95]{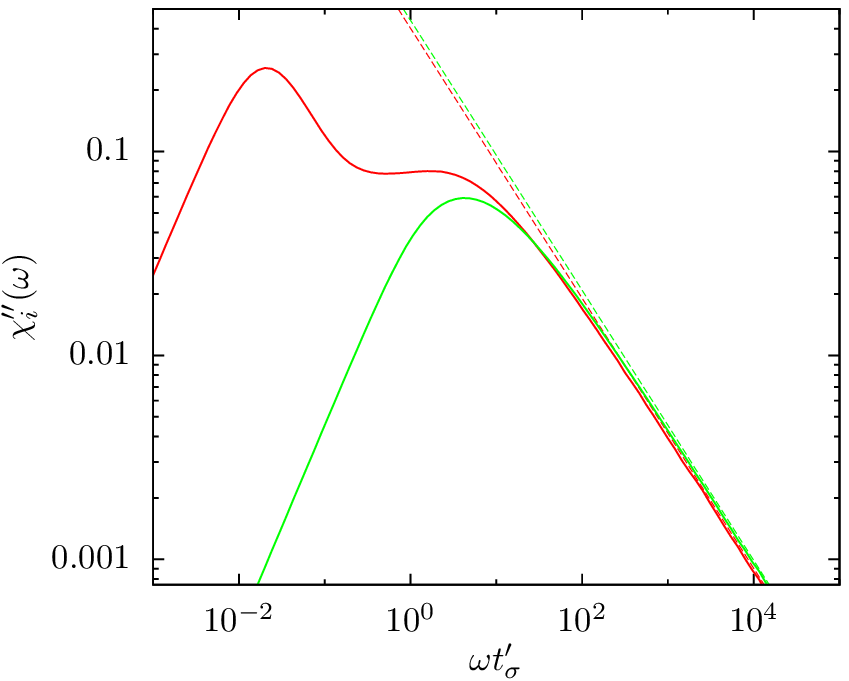}
	\caption{Susceptibility spectra $ \chi_i^{\prime \prime}(\omega) $    of the critical correlators (left) and the $ \alpha $-relaxation processes (right) for the same parameters  as in Fig.~\ref{fig:time_crit}. The asymptotes are calculated according to Eqs.~(\ref{eq:crit_asymptote_freq}) (left) and (\ref{eq:alpha_asymptote_freq}) (right).   } 
	\label{fig:susc_crit}
\end{figure}

From the correlators discussed in the last paragraphs we also calculate the susceptibility spectra \emph{via} Fourier transformation,  relying on  the modified Filon-Tuck algorithm \cite{Abramowitz19070,Tuck1967}.
 The results are presented in Fig.~\ref{fig:susc_full}. As expected from the two-step relaxation scenario two distinct peaks are observed  in the spectra for $  \epsilon < 0  $. Both peaks are described using the critical spectra derived in the previous chapter (see Fig.~\ref{fig:susc_crit}). The figure shows that for several orders of magnitude in frequency a very good overlap between the asymptotic power laws and the numerical solution of the full MCT equations holds. All of these findings are in accordance to similar observations in bulk MCT \cite{Franosch1997}. There is, however, one remarkable feature in the calculated susceptibility spectra, viz. the shoulder in the low-frequency peak in $ \chi^{\prime\prime}_1(\omega) $, that is specific for our model and that corresponds to the observed ``kink'' in the correlation function. 
A similar feature has been identified as the Cole-Cole law in Ref.~\cite{Sperl:PhysRevE.74.011503}, but in that case the ``kink'' emerges for 
frequencies larger than the von Schweidler law and thus cannot explain the shoulder observed in our model.  The small-frequency susceptibility spectrum predicted by MCT with a single decay channel can be described by an asymmetric peak. (see also Fig.~\ref{fig:susc_full}, right panel, since mode 2 has only one relaxation channel.)  The right flank of this peak is given by the von Schweidler law 
and the left side corresponds to a Debye peak.  The origin of the emergence of an additional shoulder therefore seems to be intrinsic for multiple decay channels: 
The correlator $ S_2(t) $ decays slightly faster than $ S_1(t) $ due to the smaller amplitude of $ F_2^\text{c} $. We thus  observe a faster relaxation of $ \mathcal{M}^\perp_1(t) $ compared to $ \mathcal{M}^\parallel_1(t) $ due to the asymmetry between $ \delta^{\parallel} $ and $ \delta^{\perp} $. Therefore, two overlapping peaks emerge in $ \chi^{\prime\prime}_1(\omega) $, directly connected to those two decay channels. To support this explanation we have included the corresponding Debye peaks in Fig.~\ref{fig:susc_full}, showing that the resulting spectrum, $ \chi^{\prime\prime}_1(\omega) $, is indeed emerging from two distinct relaxation processes. 

 The observation of double peaks  is an interesting finding  in itself  not directly connected to mode-coupling theory, but could also be observable in simulations or experiments for systems where multiple, very different decay channels come into play.

\section{Summary and conclusion}
\label{sec:conclusion}

An asymptotic analysis for MCT with multiple decay channels has been elaborated based on very moderate assumptions. We have found that non-trivial terms arise in the expansion in the vicinity of the glass form factors, 
leading to a slightly different form of the $ \beta $-scaling equation. Using the scale invariance of this equation, we showed that its original form and thus its universality can be recovered by a straightforward rescaling of the critical amplitude and the fundamental constant $ \lambda $. The rescaling factor $ \Delta $ is largest if there is a strong asymmetry between the different decay channels.

To demonstrate the applicability of the derived asymptotic scaling laws we have suggested a novel schematic model - a generalization of the Bosse-Krieger model - with two decay channels. As anticipated, 
the numerical solution of the MCT and the asymptotic time fractals are in very good agreement. As interesting new feature emerging already in this schematic  we have observed a pronounced shoulder in the low-frequency susceptibility spectrum of the model which we have rationalize as an indicator for multiple decay channels.

The results of this work finally enable an asymptotic analysis of MCT for systems that are characterized by multiple decay channels. Considering that most of the recently applied MCTs, viz. MCT for active particles, MCT for molecules and MCT in strongly confined systems, fall into this category, several important applications of the presented theory could come up. Furthermore, it will be very interesting to investigate whether the observed asymmetry as trace for multiple decay channels is measurable in realistic systems via simulations or experiments.

The derivation of the $\beta$-scaling law for the dynamics in the vicinity of the glass form factors relies,  first, on  a discontinuous type-B transition and, second, on the assumption that the wave vector dependence can be discretized. 
The first assumption is violated in  active microrheology where mixed continuous/discontinuous transition was observed \cite{Gruber2016,Gruber2019_2} and correspondingly a separate analysis is required. The second assumption is stricly speaking never fulfilled. For small wavenumbers hydrodynamic modes become arbitrarily slow and the structural relaxation does not dominate the dynamics of the intermediate scattering functions. Differently speaking the limit of control parameter approaching the critical value and the limit of small wavenumber do not commute. Then the hydrodynamic regime is expected to shrink as the glass transition is approached. Nevertheless the long-wavelength singularities are encoded properly in mode-coupling theory, yet  the $\beta$-scaling law cannot hold uniformly for all wavenumbers. Recently it has been shown that the interplay of structural relaxation and the hydrodynamic behavior can be addressed within mode-coupling theory~\cite{Mandal:PhysRevLett.123.168001} by using a logarithmically spaced grid of wavenumbers.

\section*{Acknowledgments}

GJ and TF gratefully acknowledge support by the Austrian Science Fund (FWF): I 2887 and TV acknowledges partial funding by DFG VO 1270/7-2.

\appendix

\section{Expansions to second order}
\label{app:exp}
Here we briefly sketch the expansion in powers of $\sqrt{|\sigma|}$ for the memory kernels and density correlation function consistently to and including order $|\sigma|$. 
 With the ansatz
\begin{equation}
 \mathbf{S}(t) = \mathbf{F}_\text{c} + \sqrt{|\sigma|} \mathbf{G}^{(1)}(t) + |\sigma| \mathbf{G}^{(2)}(t),
\end{equation}
we find for the memory kernel
\begin{align}
 \disphat{\bm{\mathcal{M}}}(z) = -\frac{1}{z} \bm{\mathcal{F}}_\text{c}+ 2 \sqrt{|\sigma| }  \bm{\mathcal{F}}_\text{c}[\hat{\mathbf{G}}^{(1)}(z), \mathbf{F}_c] -\frac{1}{z} \Delta\bm{\mathcal{F}} 
 +  2 |\sigma| \bm{\mathcal{F}}_\text{c}[\hat{\mathbf{G}}^{(2)}(z), \mathbf{F}_\text{c}] +  |\sigma| \text{LT}\{ \bm{\mathcal{F}}_\text{c}[ \mathbf{G}^{(1)}(t), \mathbf{G}^{(1)}(t)] \} ,
\end{align}
where we abbreviated the value of the MCT functional at its critical value by $\bm{\mathcal{F}}_\text{c}= \bm{\mathcal{F}}_\text{c}[\mathbf{F}_\text{c}, \mathbf{F}_\text{c}]$, and by $\Delta\bm{\mathcal{F}} = \bm{\mathcal{F}}[\mathbf{F}_\text{c}, \mathbf{F}_\text{c}] -\bm{\mathcal{F}}_\text{c}[\mathbf{F}_\text{c}, \mathbf{F}_\text{c}] = \mathcal{O}(\sigma)$ the regular variation of the MCT functional.

We will repeatedly encounter the problem of expanding the inverse of a matrix to second order in a small perturbation. This is achieved with the identity 
\begin{equation}
( \mathbf{A} + \mathbf{B} )^{-1} = \mathbf{A}^{-1} -  \mathbf{A}^{-1} \mathbf{B} ( \mathbf{A} + \mathbf{B} )^{-1} \approx \mathbf{A}^{-1} - \mathbf{A}^{-1} \mathbf{B} \mathbf{A}^{-1} + \mathbf{A}^{-1} \mathbf{B} \mathbf{A}^{-1} \mathbf{B} \mathbf{A}^{-1} .    
\end{equation}

Then the inverse of the memory kernel reads
\begin{align}
 \disphat{\bm{\mathcal{M}}}(z)^{-1} =& - z \bm{\mathcal{F}}_\text{c}^{-1} - 2 \sqrt{|\sigma|}z^2 \, \bm{\mathcal{F}}_\text{c}^{-1} \bm{\mathcal{F}}_\text{c}[\hat{\mathbf{G}}^{(1)}(z),\mathbf{F}_\text{c}]  \bm{\mathcal{F}}_\text{c}^{-1}  \nonumber \\
& + z  \bm{\mathcal{F}}_\text{c}^{-1} \Delta\bm{\mathcal{F}} \, \bm{\mathcal{F}}_\text{c}^{-1} 
-  2 |\sigma| z^2\bm{\mathcal{F}}_\text{c}^{-1} \bm{\mathcal{F}}_\text{c}[\,  \mathbf{G}^{(2)}(z) , \mathbf{F}_\text{c} ]  \bm{\mathcal{F}}_\text{c}^{-1} \nonumber \\
& -  |\sigma| z^2 \, \bm{\mathcal{F}}_\text{c}^{-1} \, \text{LT}\{ \bm{\mathcal{F}}_\text{c}[ \mathbf{G}^{(1)}(t), \mathbf{G}^{(1)}(t)] \} \bm{\mathcal{F}}_\text{c}^{-1} \nonumber \\ 
& - 4 |\sigma| z^3 \, \bm{\mathcal{F}}_\text{c}^{-1} \bm{\mathcal{F}}_\text{c}[\hat{\mathbf{G}}^{(1)}(z), \mathbf{F}_\text{c}] \bm{\mathcal{F}}_\text{c}^{-1} \bm{\mathcal{F}}_\text{c}[\hat{\mathbf{G}}^{(1)}(z), \mathbf{F}_\text{c}] \bm{\mathcal{F}}_\text{c}^{-1}.
\end{align}
The contraction $\mathcal{C}\{ \ldots \} $ of the preceeding expression yields then $-\hat{\mathbf{K}}(z)$ to the desired order. In particular from the zeroth order $\mathbf{N}_\text{c}^{-1} = \mathcal{C}\{ \bm{\mathcal{F}}_\text{c}^{-1} \} $. 

The inverse density correlation function follows from the equation of motion $\hat{\mathbf{S}}(z)^{-1} = - z \mathbf{S}^{-1} - \mathbf{S}^{-1} \hat{\mathbf{K}}(z) \mathbf{S}^{-1}$. Since the structure factor varies as $\mathbf{S} = \mathbf{S}_\text{c} + \sigma \mathbf{S}^{(1)}$, additional terms are generated in the expansion.  Collecting terms one arrives at
\begin{align}
 \hat{\mathbf{S}}(z)^{-1} =& - z \mathbf{S}_\text{c}^{-1} + \mathbf{S}_\text{c}^{-1} \mathcal{C}\{ \bm{\mathcal{M}}(z)^{-1} \} \mathbf{S}_\text{c}^{-1}  \nonumber \\
& + z \mathbf{S}_\text{c}^{-1} \sigma \mathbf{S}^{(1)} \mathbf{S}_\text{c}^{-1} + z \mathbf{S}_\text{c}^{-1} \sigma \mathbf{S}^{(1)} \mathbf{S}_\text{c}^{-1} \mathbf{N}_\text{c}^{-1} \mathbf{S}_\text{c}^{-1}
 + z \mathbf{S}_\text{c}^{-1} \mathbf{N}_\text{c}^{-1} \mathbf{S}_\text{c}^{-1} \sigma \mathbf{S}^{(1)} \mathbf{S}_\text{c}^{-1}.
\end{align}
In the contraction term a pattern is repeated which suggests to introduce the functional
\begin{align}
 \mathbf{N}_\text{c}^{(1)}[\mathbf{E}, \mathbf{F} ] := \mathbf{N}_\text{c} \mathcal{C}\{ \bm{\mathcal{F}}_c^{-1} \bm{\mathcal{F}}_\text{c}[ \mathbf{E},\mathbf{F}] \bm{\mathcal{F}}_\text{c}^{-1} \} \mathbf{N}_\text{c}.
\end{align}
Repeated use of the  equation for the glass form factor $\mathbf{F}_\text{c}^{-1} = \mathbf{S}_\text{c}^{-1} + \mathbf{S}_\text{c}^{-1} \mathbf{N}_\text{c}^{-1} \mathbf{S}_\text{c}^{-1}$ and the resulting 
identity $\mathbf{N}_\text{c} = \mathbf{S}_\text{c}^{-1} \mathbf{F}_\text{c} \tilde{\mathbf{S}}_\text{c}$ yields
\begin{align}
 \hat{\mathbf{S}}(z)^{-1} =& - z \mathbf{F}_\text{c}^{-1} + 2 \sqrt{|\sigma|} z^2 \,  \mathbf{F}_\text{c}^{-1} \tilde{\mathbf{S}}_\text{c} \mathbf{N}_\text{c}^{(1)}[\hat{\mathbf{G}}^{(1)}(z), \mathbf{F}_\text{c} ] \tilde{\mathbf{S}}_\text{c} \mathbf{F}_\text{c}^{-1} \nonumber \\
& + z \mathbf{S}_\text{c}^{-1} \sigma \mathbf{S}^{(1)} \mathbf{F}_\text{c}^{-1} + z \mathbf{F}_\text{c}^{-1} \sigma \mathbf{S}^{(1)} \mathbf{S}_\text{c}^{-1} - z \mathbf{S}_\text{c}^{-1} \sigma \mathbf{S}^{(1)} \mathbf{S}_\text{c}^{-1}  + z \mathbf{S}_c^{-1} \mathcal{C}\{ \bm{\mathcal{F}}_\text{c}^{-1} \Delta \bm{\mathcal{F}} \bm{\mathcal{F}}_\text{c}^{-1} \} \mathbf{S}_\text{c}^{-1} \nonumber \\
& - 2 |\sigma| z^2\mathbf{F}_\text{c}^{-1} \tilde{\mathbf{S}}_\text{c} \mathbf{N}_\text{c}^{(1)}[ \mathbf{G}^{(2)}(z), \mathbf{F}_\text{c} ] \tilde{\mathbf{S}}_\text{c} \mathbf{F}_\text{c}^{-1} 
- |\sigma| z^2  \, \mathbf{F}_\text{c}^{-1} \tilde{\mathbf{S}}_\text{c} \text{LT}\{  \mathbf{N}_\text{c}^{(1)}[ \mathbf{G}^{(1)}(t), \mathbf{G}^{(1)}(t)] \} \tilde{\mathbf{S}}_\text{c} \mathbf{F}_\text{c}^{-1} 
 \nonumber \\
& - 4 |\sigma| z^3  \mathbf{S}_c^{-1} \mathcal{C}\{  \bm{\mathcal{F}}_\text{c}^{-1} \bm{\mathcal{F}}_\text{c}[ \hat{\mathbf{G}}^{(1)}(z) , \mathbf{F}_\text{c} ] \bm{\mathcal{F}}_\text{c}^{-1} \bm{\mathcal{F}}_\text{c} [ \hat{\mathbf{G}}^{(1)}(z),\mathbf{F}_\text{c} ] \bm{\mathcal{F}}_\text{c}^{-1}  
\}  \mathbf{S}^{-1}_\text{c} .
\end{align}
Inverting the series once more, we find
\begin{align}\label{eq:Appendix_Sz}
\hat{\mathbf{S}}(z) =& -\frac{1}{z} \mathbf{F}_\text{c} + 2 \sqrt{|\sigma|}  \,  \tilde{\mathbf{S}}_\text{c} \mathbf{N}_\text{c}^{(1)}[\hat{\mathbf{G}}^{(1)}(z), \mathbf{F}_\text{c} ] \tilde{\mathbf{S}}_\text{c}  \nonumber \\
& - \frac{1}{z} \mathbf{F}_\text{c}  \mathbf{S}_\text{c}^{-1} \sigma \mathbf{S}^{(1)}  - \frac{1}{z} \sigma \mathbf{S}^{(1)}    \mathbf{S}_\text{c}^{-1} \mathbf{F}_\text{c} + \frac{1}{z} \mathbf{F}_\text{c} \mathbf{S}_\text{c}^{-1} \sigma \mathbf{S}^{(1)} \mathbf{S}_\text{c}^{-1} \mathbf{F}_\text{c} 
- \frac{1}{z} \mathbf{F}_\text{c} \mathbf{S}_\text{c}^{-1} \mathcal{C} \{ \bm{\mathcal{F}}_\text{c}^{-1} \Delta\bm{\mathcal{F}} \bm{\mathcal{F}}_\text{c}^{-1} \} \mathbf{S}_c^{-1} \mathbf{F}_\text{c} \nonumber \\
& + 2 |\sigma| \tilde{\mathbf{S}}_\text{c} \mathbf{N}_\text{c}^{(1)}[  \mathbf{G}^{(2)}(z) , \mathbf{F}_\text{c} ] \tilde{\mathbf{S}}_c 
+ |\sigma|  \tilde{\mathbf{S}}_c \text{LT}\{  \mathbf{N}_\text{c}^{(1)}[ \mathbf{G}^{(1)}(t), \mathbf{G}^{(1)}(t)] \} \tilde{\mathbf{S}}_\text{c} \nonumber  \\
& + 4 |\sigma| z \mathbf{F}_\text{c}{\mathbf{S}}_\text{c}^{-1} \mathcal{C} \{ \bm{\mathcal{F}}_\text{c}^{-1}  \bm{\mathcal{F}}_\text{c}[\hat{\mathbf{G}}^{(1)}(z), \mathbf{F}_\text{c} ] \bm{\mathcal{F}}_\text{c}^{-1}  \bm{\mathcal{F}}_\text{c}[\hat{\mathbf{G}}^{(1)}(z), \mathbf{F}_\text{c} ] \bm{\mathcal{F}}_\text{c}^{-1}  \}    {\mathbf{S}}_\text{c}^{-1}  \mathbf{F}_\text{c} \nonumber \\
& - 4 |\sigma| z  \tilde{\mathbf{S}}_\text{c} \mathbf{N}_\text{c}^{(1)}[\hat{\mathbf{G}}^{(1)}(z), \mathbf{F}_\text{c} ] \tilde{\mathbf{S}}_\text{c} \mathbf{F}_\text{c}^{-1}  \tilde{\mathbf{S}}_\text{c} \mathbf{N}_\text{c}^{(1)}[ \hat{\mathbf{G}}^{(1)}(z), \mathbf{F}_\text{c} ] \tilde{\mathbf{S}}_\text{c}.
\end{align}
Here, all terms in the second line can be combined to find a compact term that also includes all linear terms in $ \sigma $,
\begin{align}\label{eq:Appendix_regular}
- &\frac{1}{z} \mathbf{F}_\text{c}  \mathbf{S}_\text{c}^{-1} \sigma \mathbf{S}^{(1)}  - \frac{1}{z} \sigma \mathbf{S}^{(1)}    \mathbf{S}_\text{c}^{-1} \mathbf{F}_\text{c} + \frac{1}{z} \mathbf{F}_\text{c} \mathbf{S}_\text{c}^{-1} \sigma \mathbf{S}^{(1)} \mathbf{S}_\text{c}^{-1} \mathbf{F}_\text{c} 
- \frac{1}{z} \mathbf{F}_\text{c} \mathbf{S}_\text{c}^{-1} \mathcal{C} \{ \bm{\mathcal{F}}_\text{c}^{-1} \Delta\bm{\mathcal{F}} \bm{\mathcal{F}}_\text{c}^{-1} \} \mathbf{S}_c^{-1} \mathbf{F}_\text{c} \nonumber \\
=& \tilde{\mathbf{S}}_\text{c}\mathbf{S}_\text{c}^{-1} \left(  \mathbf{S} \mathbf{N}_\epsilon(\mathbf{S}-\mathbf{F}_\text{c}) -\mathbf{S}_\text{c} \mathbf{N}_\text{c}\tilde{\mathbf{S}}_\text{c}   \right),
\end{align}
as can be seen from a straightforward expansion of the right-hand side. Collecting the terms in Eq.~\eqref{eq:Appendix_Sz} to orders $\mathcal{O}(|\sigma|^0),\mathcal{O}(|\sigma|^{1/2}), \mathcal{O}(|\sigma|)$ yields the zeroth, first, and second order expansion of Sec.~\ref{Sec:Asymptotic_Expansion}.

\section{Numerical solution of the schematic model}
\label{ap:numerics}

Introducing the effective memory kernel in a similar way as was done in Ch.~\ref{sec:analysis} we obtain,
\begin{eqnarray}
\ddot{S}_i(t) + QS_i(t) &+& Q\int_0^t M_i(t-t')\dot{S}_i(t')\text{d}t'=0, \label{eq:schematic_start}\\
\dot{M}_i(t) +  \int_0^t \alpha_i(t-t') M_i(t') \text{d}t' &=& \dot{\beta}_i(t) +\frac{Q^\perp Q^\parallel}{Q} \int_0^t \tilde{\mathcal{M}}^\parallel_i(t-t') \tilde{\mathcal{M}}^\perp_i(t') \text{d}t',  \label{eq:schematic_eff}\\
\alpha_i(t) &=& \frac{Q^\perp Q^\parallel}{Q} (\tilde{\mathcal{M}}_i^\parallel(t)+\tilde{\mathcal{M}}_i^\perp(t)),\\
\beta_i(t) &=& \frac{{Q^\parallel}^2}{Q^2} \tilde{\mathcal{M}}_i^\parallel(t)+\frac{{Q^\perp}^2}{Q^2} \tilde{\mathcal{M}}_i^\perp(t)\label{eq:schematic_final},
\end{eqnarray}
with $ i =1,2 $ and $ Q = Q^\perp + Q^\parallel. $   The instantaneous contributions to the memory kernels, $ \nu_i $, which were not considered in the above derivation of the effective memory kernel, will be directly included to $ \bm{ \mathcal{\tilde{M}}}(t=0) $ after discretization such that,
	\begin{align}
\mathcal{\tilde{M}}_1^\alpha(q,t=0) &=  k\left[\delta^\alpha S_1(t)^2+(1-\delta^\alpha)S_2(t)^2\right]  + \nu_1/\Delta t,\\
\mathcal{\tilde{M}}_2^\alpha(q,t=0)&=  krS_1(t)S_2(t)  + \nu_2/\Delta t.
\end{align}

The equations of motion for the correlation functions $ S_i(t) $ in the schematic model (\ref{eq:schematic_start}) are similar to the ones found for MCT in bulk systems and we have applied the same discretization scheme as suggested in Ref.~[\citen{Sperl2000}]. The effective memory kernel can then be integrated using the techniques suggested in Ref.~[\citen{Gruber2019_2}] for  this type of differential equation (see chaper 3.5.1. ``Integral method with moments'').

The above described discretization route yields good results for the short-time behavior of the correlation functions, however, for long times, the emergence of instabilities is observed. We therefore had to utilize several techniques to stabilize the numerical scheme, which we will list in the following.
\begin{itemize}
	\item Due to cancellation effects, the integral method with moments becomes unstable for long times. We therefore derived the corresponding integro-differential equation by taking the time derivative on both sides of Eq.~(\ref{eq:schematic_eff}) and solved it using the method described in Ref.~[\citen{Gruber2019_2}] (see chaper 3.5.1. ``Integro-differential method with moments'').
	\item In the applied schemes, the time derivative of the convolution integral,
	\begin{equation}\label{key}
	\frac{\text{d}}{\text{d}t} \int_0^t A(t-t') B(t') \text{d}t',
	\end{equation}
	for time-dependent functions $ A(t),B(t) $ is always discretized as,
	\begin{equation}\label{eq:discrte_asym}
	 A_{\bar{i}} {B}_{i-\bar{i}}   + \sum_{j=1}^{i- \bar{i}} ( A_{i-j+1}- A_{i-j} ) dB_j + \sum_{j=1}^{\bar{i}}  dA_j ( B_{i-j+1}- B_{i-j}  ), 
	\end{equation}
	with $ A_i = A(t_i) $, $ t_i = i\Delta_t $, moments $ \text{d}A_i = \Delta_t^{-1} \int_{t_{i-1}}^{t_i}A(t')\text{d}t' $ and $ \bar{i}=\lfloor i/2  \rfloor  $. Here, the brackets $ \lfloor j \rfloor $ denote the largest integer less or equal $ j $. For odd $ i $ we thus introduce an asymmetry between the functions $ A(t) $ and $ B(t) $ which is obviously an artifact of the discretization. We resolved this by a trivial rewriting of Eq.~(\ref{eq:discrte_asym}),
	\begin{align}\label{eq:discrte_sym}
	 \frac{1}{2} \left(  A_{\bar{i}} {B}_{i-\bar{i}} +A_{i-\bar{i}} {B}_{\bar{i}} \right)    &+ \frac{1}{2}\sum_{j=1}^{i- \bar{i}} ( A_{i-j+1}- A_{i-j} ) dB_j + \frac{1}{2}\sum_{j=1}^{\bar{i}}  dA_j ( B_{i-j+1}- B_{i-j}  ),\nonumber\\
	 &+\frac{1}{2}\sum_{j=1}^{\bar{i}} ( A_{i-j+1}- A_{i-j} ) dB_j + \frac{1}{2}\sum_{j=1}^{i-\bar{i}}  dA_j ( B_{i-j+1}- B_{i-j}  ).
	\end{align}
\end{itemize}

\bibliography{library,library_local,library_local}

\end{document}